\documentclass[fleqn,usenatbib]{mnras}

\usepackage{newtxtext,newtxmath}
\usepackage{booktabs} % added by author
\usepackage[T1]{fontenc}
\usepackage{ae,aecompl}

\usepackage{longtable}

\usepackage{graphicx}	% Including figure files
\usepackage{amsmath}	% Advanced maths commands
\usepackage{amssymb}	% Extra maths symbols
\title[LMEBs in the WTS: the persistence of the M-dwarf radius inflation problem]{Low-mass eclipsing binaries in the WFCAM Transit Survey: the persistence of the M-dwarf radius inflation problem}

\author[P. Cruz et al.]{
Patricia Cruz,$^{1}$\thanks{E-mail: patricia.cruz@usp.br}
Marcos Diaz,$^{1}$
Jayne Birkby,$^{2}$
David Barrado,$^{3}$
Brigitta Sip\"ocz,$^{4}$
\newauthor
 and Simon Hodgkin$^{4}$\\
\\
$^{1}$Departamento de Astronomia, Instituto de Astronomia, Geof\'isica e Ci\^encias Atmosf\'ericas, Universidade de S\~ao Paulo (IAG/USP),\\
Rua do Mat\~ao 1226, 05508-900, S\~ao Paulo, Brazil\\
$^{2}$Anton Pannekoek Institute for Astronomy, University of Amsterdam, Science Park 904, 1098 XH Amsterdam, The Netherlands\\
$^{3}$Depto. de Astrof\'isica, Centro de Astrobiolog\'ia (INTA-CSIC), ESAC campus, P.O. Box 78, E-28691 Villanueva de la Ca\~nada, Spain\\
$^{4}$Institute of Astronomy, University of Cambridge, Madingley Road, Cambridge, CB3 0HA, UK
}

\date{Accepted XXX. Received YYY; in original form ZZZ}

\pubyear{2017}

\begin{document}
\label{firstpage}
\pagerange{\pageref{firstpage}--\pageref{lastpage}}
\maketitle

\begin{abstract}
We present the characterization of 5 new short-period low-mass eclipsing binaries from the WFCAM Transit Survey. The analysis was performed by using the photometric WFCAM J-mag data and additional low- and intermediate-resolution spectroscopic data to obtain both orbital and physical properties of the studied sample. 
The light curves and the measured radial velocity curves were modeled simultaneously with the JKTEBOP code, with Markov chain Monte Carlo simulations for the error estimates. 
The best-model fit have revealed that the investigated detached binaries are in very close orbits, with orbital separations of $2.9 \leq a \leq 6.7$ $R_{\odot}$ and short periods of $0.59 \leq P_{\rm orb} \leq 1.72$ d, approximately. We have derived stellar masses between $0.24$ and $0.72$ $M_{\odot}$ and radii ranging from $0.42$ to $0.67$ $R_{\odot}$. 
The great majority of the LMEBs in our sample has an estimated radius far from the predicted values according to evolutionary models. The components with derived masses of $M < 0.6$ $M_{\odot}$ present a radius inflation of $\sim$$9\%$ or more. This general behavior follows the trend of inflation for partially-radiative stars proposed previously. 
These systems add to the increasing sample of low-mass stellar radii that are not well-reproduced by stellar models. They further highlight the need to understand the magnetic activity and physical state of small stars. 
Missions like TESS will provide many such systems to perform high-precision radius measurements to tightly constrain low-mass stellar evolution models.

\end{abstract}


%%%%%%%%%%%%%%%%%%%%%%%%%%%%%%%%%%%%%%%%%%%%%%%%%%

%%%%%%%%%%%%%%%%% BODY OF PAPER %%%%%%%%%%%%%%%%%%

\section{Introduction}

Approximately half of the stars in our galaxy are part of binary systems \citep{Duquennoy91,Raghavan10} and that these systems are expected to be formed at the pre-main sequence phase. Binary systems can be formed, for instance, by the fragmentation of proto-stellar disks, where they are gravitationally bound before the disks are %swept away
dissipated by stellar winds \citep[][and references therein]{Tobin13}. However, several formation and evolutionary processes are still not completely understood, as for example, the loss of angular momentum that results in the migration of these binaries to closer orbits, with short periods of just a few days \citep{Nefs12}.

Binary systems can be separated in three groups -- detached, semi-detached, and contact binaries -- that can be interpreted as different evolutionary stages. As the most massive component of a close binary system evolves, a mass transfer episode starts and the system goes from a detached to a semi-detached stage. If this mass transfer happens in a very high rate, both binary components share the same envelope and become a contact binary. At the end of this process, the system has a smaller orbital separation and is composed by a main-sequence (MS) and an evolved star, starting a second detached phase. Further binary angular momentum losses may lead to Roche lobe contact of the less massive component which starts to transfer mass to the evolved component, a white dwarf for example, and starts a second semi-detached phase. This system is known as a Cataclysmic Variable (CV), and define a class of semi-detached binaries of short period where the secondary component is typically a low-mass star. Then, the evolution of current short-period close binary systems can be associated with phenomena like CVs and Type Ia Supernovae. 

%As argued by \citet{Andersen91}, 
Eclipsing binaries (EBs), and particularly the detached spectroscopic double-lined systems, give the most precise ways to derive the physical parameters of low-mass stars without the use of stellar models \citep[][and references therein]{Andersen91,Torres10}. However, only a small number (a few tens) of well-characterized low-mass eclipsing binaries (LMEBs) can be found in the literature \citep{Southworth15}. These known systems present an intriguing trend when compared to stellar models: the measured stellar radii are usually 5 to 10\% bigger than the expected value \citep[e.g., ][]{Lopez05,Kraus11,Birkby12,Nefs13,Dittmann17}. This behavior is known as the radius anomaly of low-mass stars and it is a recurrent problem that should not be neglected.

A significant radius anomaly has also been observed in low-mass stars that are synchronous members of semi-detached binaries. This phenomenon is often attributed to the significant companion mass loss by its Roche lobe overflow \citep{Knigge11}. In this case, the thermal and mass-loss time-scales of the companion star may be comparable, leading to a slight deviation from thermal equilibrium and larger radii. 

A few theories have been appeared on the literature as attempts of explaining the radius anomaly in detached binary systems. One that is largely discussed considers the magnetic activity as a possible reason for the radius inflation \citep{Torres06,Chabrier07}. The majority of well-characterized M-dwarf eclipsing binaries are in very close systems, with periods of less than 2 days. These short-period systems should be synchronized and in circular orbits due to tidal effects \citep{Zahn77}, which would enhance the magnetic activity as a consequence \citep{Chabrier07,Kraus11,Birkby12}. This activity enhancement hypothesis is sustained by observations of H${\rm \alpha}$ and X-ray emission from one or both of the binary components \citep[e.g., ][]{Chabrier07,Huelamo09,Kraus11}.

It is unquestionable the need for a greater sample of well-characterized LMEBs to study the radius inflation and its effect on the evolution of low-mass stars as components of detached and semi-detached binary systems. Thus, the discovery and the characterization of every single new LMEB is significantly important to help understand the radius inflation issue. We, then, present here the characterization of 5 new low-mass systems from the WFCAM Transit Survey.

\section{The WFCAM Transit Survey}
\label{sec:photobs}

The WFCAM Transit Survey \citep[WTS, ][]{Birkby11,Kovacs13} was a photometric long-term back-up program running on the 3.8-m United Kingdom Infrared Telescope (UKIRT) at Mauna Kea, Hawaii. This survey had two main aims: the search for exoplanets around cool stars and the detection of a great number of low-mass eclipsing binaries. The survey objectives and prospects were already presented and described in detail by \citet{Birkby11,Birkby12,Kovacs13}, thus, only a brief summary is presented in this section.

\subsection{The UKIRT/WFCAM photometry}

The photometric data was acquired with the Wide-Field Camera \citep[WFCAM, ][]{Hodgkin09}, in the J band ($1.25$ $\mu$m), close to the maximum of the spectral energy distribution (SED) of low-mass stars. Individual images were also obtained with the five near-infrared WFCAM filters (Z, Y, J, H, K) for the characterization of the observed objects.
The survey observed four different regions of the sky centered at Right Ascension (RA) hours of 03, 07, 17 and 19h, providing targets for all year. Each field covered 1.5 degrees of the sky every 15 min (9-point jitter pattern $\times$ 10s exposure time $\times$ 8 pointings $+$ overheads), for a cadence of 4 points per hour.

The data processing and the generation of light curves (LC) followed the procedures described by \citet{Irwin09}. Succinctly, the image processing was performed by the Cambridge Astronomical Survey Unit pipeline to remove instrumental signature for infrared arrays, dark and reset anomaly. This pipeline also performs flat-fielding, decurtaining and sky subtraction. The photometric calibration was done by using cataloged point-sources present in the same frame and according to \citet{Hodgkin09}.

Additional details on the image reduction and light curve generation procedures are deeply reported in \citet[][and references therein]{Kovacs13}.

\subsection{Eclipse detection and candidate selection}

An automated search for eclipses was performed in all generated light curves. For that, the Box-Least-Squares (BLS) algorithm \citep[OCCFIT, ][]{Aigrain04} was chosen as the simplest approach to look for planetary transit-like events (the main goal of the survey). The code searches for a periodic dim in the light curve, compared to the mean stellar flux, and designs a box-shaped eclipse adopting a single-bin method. This procedure considers all in-occultation data points as a single bin and generates a box-shaped eclipse-like event. This simplistic method is efficient for detection purposes \citep{Birkby12,Birkby14} and the significance of the eclipses detected with OCCFIT was already discussed by \citet[][and references therein]{Miller08}.

As performed previously by \citet{Birkby12} for the WTS 19hr field, we ran OCCFIT on the light curves of the 17hr field as well. The OCCFIT detection statistic {$S$} \citep[see ][for more details]{Pont06}, which assesses the significance of the detections, was defined to be always $S$ $\geq 5$ and the derived orbital period should be outside the interval $0.99 > P_{\rm orb} > 1.005$ days, a common window-function alias of the WFCAM Transit Survey.

The detailed description of the eclipse detection procedure and the EB candidates selection can be found in \citet{Birkby12,Kovacs13}.

\begin{table}%[!h]
\begin{center}
\caption{Light curves for the 5 LMEBs from the WFCAM Transit Survey reported in this paper. (This table is fully published online.)}
\label{photAllEBs}
\begin{tabular}{cccc}
\hline \hline \\ [-3ex]
Binary & MHJD$^{ \rm *}$ & J$_{\rm WFCAM}$ & $\sigma_{\rm WFCAM}$ \\
Name  & (d) & (mag) & (mag)  \\

%[0.5ex] 
\hline\\  [-2ex]

17e-3-02003 & 54552.58495455 & 15.2560 & 0.0060 \\ 
17e-3-02003 & 54552.59788564 & 15.2566 & 0.0073 \\ 
17e-3-02003 & 54552.60936660 & 15.2586 & 0.0067 \\ 
17e-3-02003 & 54552.62178764 & 15.2498 & 0.0063 \\ 
17e-3-02003 & 54561.60691676 & 15.2545 & 0.0056 \\ 
17e-3-02003 & 54561.61840766 & 15.2510 & 0.0058 \\ 
17e-3-02003 & 54561.62972854 & 15.2637 & 0.0058 \\ 
17e-3-02003 & 54561.64101942 & 15.2596 & 0.0060 \\ 
%17e-3-02003 & 54568.58839266 & 15.2641 & 0.0061 \\ 
%17e-3-02003 & 54568.59961347 & 15.2480 & 0.0060 \\ 
%17e-3-02003 & 54568.62611538 & 15.2579 & 0.0060 \\ 
%17e-3-02003 & 54568.64771695 & 15.2568 & 0.0090 \\ 
... & ... & ... & ... \\
\\  [-1ex]
%\hline\\  [-2ex]

17h-4-01429 & 54552.59361278 & 15.8130 & 0.0102 \\
17h-4-01429 & 54552.60507375 & 15.8056 & 0.0098 \\
17h-4-01429 & 54552.61752480 & 15.7893 & 0.0096 \\
17h-4-01429 & 54552.62876574 & 15.8100 & 0.0101 \\
17h-4-01429 & 54561.61413894 & 15.8180 & 0.0085 \\
17h-4-01429 & 54561.62545983 & 15.8257 & 0.0087 \\
17h-4-01429 & 54561.63675072 & 15.8200 & 0.0085 \\
17h-4-01429 & 54561.64797160 & 15.8244 & 0.0083 \\
%17h-4-01429 & 54568.59534893 & 15.8131 & 0.0091 \\
%17h-4-01429 & 54568.61686050 & 15.8106 & 0.0093 \\
%17h-4-01429 & 54568.64334243 & 15.8074 & 0.0094 \\
%17h-4-01429 & 54568.65628337 & 15.8218 & 0.0150 \\
... & ... & ... & ... \\
\\  [-1ex]
%\hline\\  [-2ex]

19c-3-08647 & 54317.31101480 & 14.7889 & 0.0048 \\
19c-3-08647 & 54317.32307474 & 14.7915 & 0.0046 \\
19c-3-08647 & 54317.33442468 & 14.7937 & 0.0047 \\
19c-3-08647 & 54317.34567462 & 14.7973 & 0.0048 \\
19c-3-08647 & 54317.35998455 & 14.7940 & 0.0048 \\
19c-3-08647 & 54317.37133449 & 14.7874 & 0.0046 \\
19c-3-08647 & 54317.38281444 & 14.7761 & 0.0046 \\
19c-3-08647 & 54317.39393439 & 14.7986 & 0.0047 \\
%19c-3-08647 & 54319.38433236 & 14.7865 & 0.0044 \\
%19c-3-08647 & 54319.39543228 & 14.7858 & 0.0045 \\
%19c-3-08647 & 54319.40670221 & 14.7856 & 0.0044 \\
%19c-3-08647 & 54319.41802213 & 14.7884 & 0.0044 \\
... & ... & ... & ... \\
\\  [-1ex]
%\hline\\  [-2ex]

19f-4-05194 & 54628.50806538 & 515.9730 & 50.0097 \\
19f-4-05194 & 54628.52266592 & 515.9825 & 50.0097 \\
19f-4-05194 & 54628.53653644 & 515.9773 & 50.0097 \\
19f-4-05194 & 54628.54783686 & 515.9798 & 50.0094 \\
19f-4-05194 & 54628.56008731 & 516.0452 & 50.0098 \\
19f-4-05194 & 54628.57152774 & 516.1470 & 50.0109 \\
19f-4-05194 & 54628.58606828 & 516.2154 & 50.0111 \\
19f-4-05194 & 54628.59773871 & 516.1351 & 50.0106 \\
%19f-4-05194 & 54629.49441174 & 516.0193 & 50.0113 \\
%19f-4-05194 & 54629.50605217 & 515.9477 & 50.0109 \\
%19f-4-05194 & 54629.51726258 & 515.9463 & 50.0102 \\
%19f-4-05194 & 54629.52875300 & 515.9702 & 50.0107 \\
... & ... & ... & ... \\
\\  [-1ex]
%\hline\\  [-2ex]

19g-2-08064 & 54317.30968224 & 14.5019 & 0.0041 \\
19g-2-08064 & 54317.32174217 & 14.4965 & 0.0041 \\
19g-2-08064 & 54317.33308211 & 14.4966 & 0.0042 \\
19g-2-08064 & 54317.34434204 & 14.4919 & 0.0041 \\
19g-2-08064 & 54317.35865196 & 14.4889 & 0.0040 \\
19g-2-08064 & 54317.36975190 & 14.4843 & 0.0040 \\
19g-2-08064 & 54317.38149183 & 14.4875 & 0.0040 \\
19g-2-08064 & 54317.39259177 & 14.4824 & 0.0041 \\
%19g-2-08064 & 54319.38299854 & 14.4764 & 0.0038 \\
%19g-2-08064 & 54319.39408846 & 14.4716 & 0.0038 \\
%19g-2-08064 & 54319.40536837 & 14.4763 & 0.0038 \\
%19g-2-08064 & 54319.41669828 & 14.4758 & 0.0039 \\
... & ... & ... & ... \\

\hline %\hline
\end{tabular}
\begin{list}{}{}
\item[]{\scriptsize{ {\bf Note.} $^{\rm *}$ ${\rm MHJD}={\rm HJD}-2400000.5$.}}
\end{list}
\end{center}
\end{table}

Within the whole list of EB candidates found in the WTS 17hr and 19hr fields, targets were selected for a spectroscopic follow up on the basis of two main criteria. Firstly, the shape of the light curve should clearly show a detached EB system, in order to exclude interacting binaries. Secondly, effective temperature estimates from the spectral energy distribution obtained with the available broad-band photometry (described later on sect. \ref{sect:sed}) should be compatible with late-K and M dwarfs.

Among the candidates selected through these criteria, we have chosen 5 EB candidates for spectroscopic characterization, being 2 candidates from the 17hr field and 3 from the 19hr field. The objects are: 17e-3-02003, 17h-4-01429, 19c-3-08647, 19f-4-05194, and 19g-2-08064. 
We maintained the naming system described in \citet{Birkby12}, which is based on the RA of the object, the correspondent pointing, the WFCAM chip, and the target's sequence number in the WTS master catalog.

The WTS light curves were folded on the preliminary found period and a very few outliers could be found. To identify and clean the LC from these outliers, we firstly excluded the primary and secondary eclipses in order to keep only the light-curve baseline. Then, all non-consecutive data points outside a three sigma boundary from the median value of the LC baseline were considered as outliers and were discarded. %The final light curves are presented in table \ref{photAllEBs}\footnote{A complete version of the mentioned table is available online.}.
The final light curves are shortly presented in table \ref{photAllEBs} (a complete version of this table is available online).

\section{Low- and intermediate-resolution spectroscopy}
\label{sec:specobs}

\subsection{Data acquisition}

We have dedicated a total of 10 nights at the Calar Alto Observatory (CAHA), in Spain, distributed in four observing runs between July 2011 and July 2012, to spectroscopically classify and solve the selected 5 low-mass EB systems. The data were acquired with the Cassegrain TWIN Spectrograph mounted on the 3.5-m telescope. We used the red arm only because 
we did not expect much flux at shorter wavelengths, since the targets were supposed to be late-type stars.

The great majority of the observing time was dedicated to measure radial velocity (RV) shifts by taking intermediate-resolution spectra, with a dispersion of $\sim$$0.39$ \AA/pix (R$\sim$$10000$), with the grating T10 and a $1.2\arcsec$ slit. The spectral coverage of around $800$ \AA$ $ was chosen to be centered at H${\rm \alpha}$, spanning from $6200$ to $7000$ \AA. We also obtained low-resolution data to spectroscopically classify the components of the systems, with the grating T07. These spectra cover a wavelength range of around $2100$ \AA, going approximately from $6000$ to $8100$ \AA, with a dispersion of $\sim$$1.62$ \AA/pix (R$\sim$$2300$). This range comprehends several spectral features, such as molecular bands, which are crucial for performing further spectral typing.

%The exposure time and other observational details can be found in tables \ref{obs17h} and \ref{obs19h} for the objects from the 17hr and 19hr fields, respectively. The signal-to-noise ratio (SNR) shown in column 6 of each table was measured at the continuum.
The exposure time and other observational details can be found in columns 2 to 5 of table \ref{obs17-19h} 
for the objects from the 17hr and 19hr fields. Arc images were also acquired before and after each target observation for the wavelength calibration.

The data reduction was performed by using the IRAF\footnote{IRAF is distributed by the National Optical Astronomy Observatory, operated by the Association of Universities for Research in Astronomy, Inc., under cooperative agreement with the National Science Foundation.} software package following a standard procedure for CCD processing and spectrum extraction.

\begin{center}
\onecolumn
\LTcapwidth=\linewidth
\begin{longtable}{lccccccc}
\caption{Summary of the spectroscopic observations and measured radial velocities for all 5 LMEBs from the WTS. Gratings T07 and T10 represent the low- and intermediate-resolution spectra, with resolutions of approximately R$\sim$2300 and R$\sim$10000, respectively. Columns 2 to 5 present the observational details of each acquired spectrum. For the intermediate resolution spectra, the orbital phase of the observation is shown in column 6, and the respective measured radial velocities (RV) for the primary and the secondary components are shown in columns 7 and 8.}\label{obs17-19h}
\\
\hline \hline \\ [-1ex]
Object & MHJD & Grating & Exp. time & Airmass & Phase & RV$_{1}$ & RV$_{2}$  \\
Name & ({\scriptsize HJD}{\footnotesize -2400000.5}) & Number & (s) &  &  & (km s$^{-1}$) & (km s$^{-1}$)  \\
%[0.5ex] 
\hline\\  [-1ex]
\endfirsthead
\caption{continued.}\\
\hline \hline \\ [-1ex]
Object & MHJD & Grating & Exp. time & Airmass & Phase & RV$_{1}$ & RV$_{2}$  \\
Name & ({\scriptsize HJD}{\footnotesize -2400000.5}) & Number & (s) &  &  & (km s$^{-1}$) & (km s$^{-1}$)  \\
\hline\\  [-1ex]
\endhead
\hline\\  [-1ex]
\endfoot

%\multicolumn{8}{l}{\scriptsize 17HR FIELD}  \\
17e-3-02003	& 55761.93893508 & T10 & 1200  & 1.23  & 0.7848 & 115.7 & -75.2 \\ %2.91  - 20110719
			& 55762.06949593 & T10 & 1200  & 2.37  & 0.8914 & 81.3 & -54.4 \\ %2.84  - 20110719
			& 55762.92592164 & T10 & 1200  & 1.22  & 0.5903 & 75.5 & -42.8 \\ %3.53  - 20110720
			& 55762.94045863 & T10 & 1200  & 1.24  & 0.6022 & 80.9 & -49.1 \\ %4.29  - 20110720
			& 56090.91266077 & T10 & 1000  & 1.38  & 0.3304 & -55.1 & 122.9 \\ %4.01  - 20120612
			& 56090.92488191 & T10 & 1000  & 1.33  & 0.3404 & -46.6 & 120.2 \\ %3.27  - 20120612
			& 56090.93710247 & T10 & 1000  & 1.29  & 0.3504 & -44.8 & 101.3 \\ %4.54  - 20120612
			& 56090.95029776 & T10 & 1000  & 1.25  & 0.3611 & -54.5 & 93.6 \\ %5.31  - 20120612
			& 56090.96251949 & T10 & 1000  & 1.23  & 0.3711 & -43.8 & 95.2 \\ %3.98  - 20120612
			& 56090.97474642 & T10 & 1000  & 1.21  & 0.3811 & -38.1 & 97.1 \\ %3.18  - 20120612
			& 56091.93748988 & T10 & 1100  & 1.28  & 0.1674 & -57.7 & 122.7 \\ %4.02  - 20120613
			& 56091.95086866 & T10 & 1100  & 1.24  & 0.1784 & -54.2 & 121.7 \\ %3.88  - 20120613
			& 56091.96425265 & T10 & 1100  & 1.22  & 0.1893 & -79.1 & 128.3 \\ %5.79  - 20120613
			& 56092.02624221 & T10 & 1100  & 1.22  & 0.2399 & -71.4 & 129.6 \\ %5.23  - 20120613
			& 56092.03962851 & T10 & 1100  & 1.24  & 0.2508 & -68.0 & 121.8 \\ %2.62  - 20120613
			& 56092.05300728 & T10 & 1100  & 1.28  & 0.2617 & -72.9 & 123.8 \\ %6.42  - 20120613
			& 56114.94131237 & T07 & 1800  & 1.20  & -- & -- & -- \\ %12.76  - 20120706
\\  [-2ex]
17h-4-01429	& 55761.95640436 & T10 & 1000  & 1.27  & 0.7931 & -11.3 & -206.4 \\ %1.69  - 20110719
			& 55761.96863360 & T10 & 1000  & 1.31  & 0.8015 & -10.5 & -173.3 \\ %1.09  - 20110719
%			& 55762.02662469 & T10 & 1000  & 1.67  &  &  &  \\ %0.76  - 20110719
			& 55762.03884468 & T10 & 1000  & 1.80  & 0.8501 & -21.6 & -176.9 \\ %0.94  - 20110719
			& 55762.89122851 & T10 & 1000  & 1.20  & 0.4400 & -112.9 & -47.1 \\ %1.63  - 20110720
			& 55762.90346065 & T10 & 1000  & 1.20  & 0.4485 & -121.7 & -37.0 \\ %1.20  - 20110720
			& 56088.88960980 & T10 & 1200  & 1.59  & 0.1115 & -142.7 & -28.9 \\ %1.42  - 20120610
			& 56088.90415779 & T10 & 1200  & 1.48  & 0.1216 & -133.0 & -4.1 \\ %0.79  - 20120610
			& 56088.91871272 & T10 & 1200  & 1.39  & 0.1317 & -144.1 & -15.4 \\ %0.71  - 20120610
			& 56089.02424900 & T10 & 1100  & 1.21  & 0.2043 & -175.8 & 10.1 \\ %0.67  - 20120610
			& 56089.03763298 & T10 & 1100  & 1.22  & 0.2136 & -169.6 & 8.2 \\ %0.84  - 20120610
			& 56089.05104475 & T10 & 1100  & 1.25  & 0.2229 & -167.2 & 2.0 \\ %0.96  - 20120610
			& 56089.11092395 & T10 & 1200  & 1.50  & 0.2647 & -157.6 & 6.5 \\ %1.29  - 20120610
			& 56089.12547309 & T10 & 1200  & 1.61  & 0.2748 & -167.9 & 8.8 \\ %0.98  - 20120610
			& 56091.06888000 & T10 & 1000  & 1.31  & 0.6196 & -31.3 & -- \\ %1.14  - 20120612
			& 56091.08110577 & T10 & 1000  & 1.36  & 0.6281 & -26.8 & -157.2 \\ %0.98  - 20120612
			& 56091.09332576 & T10 & 1000  & 1.43  & 0.6366 & -27.3 & -179.1 \\ %0.83  - 20120612
			& 56091.10670222 & T10 & 1200  & 1.51  & 0.6466 & -20.9 & -170.4 \\ %0.93  - 20120612
			& 56091.12133008 & T10 & 1200  & 1.63  & 0.6567 & -7.8 & -175.6 \\ %0.99  - 20120612
			& 56091.13587749 & T10 & 1200  & 1.78  & 0.6668 & -23.5 & -176.5 \\ %0.70  - 20120612
			& 56091.89204349 & T10 & 1100  & 1.50  & 0.1899 & -169.7 & 5.0 \\ %0.81  - 20120613
			& 56091.90542227 & T10 & 1100  & 1.42  & 0.1991 & -184.6 & 10.1 \\ %0.83  - 20120613
			& 56091.91880394 & T10 & 1100  & 1.35  & 0.2084 & -167.2 & 14.7 \\ %1.02  - 20120613
			& 56114.96730472 & T07 & 2100  & 1.22  & -- & -- & -- \\ %5.09  - 20120706
\\  [-2ex]
%\multicolumn{8}{l}{\scriptsize 19HR FIELD}  \\
19c-3-08647	& 55782.89011630 & T10 & 700  & 1.03  & 0.7069 & 96.1 & -93.8 \\ %5.12  - 20110809
			& 55782.89880434 & T10 & 700  & 1.02  & 0.7170 & 83.7 & -95.2 \\ %3.66  - 20110809
			& 55782.91765525 & T10 & 700  & 1.01  & 0.7387 & 97.3 & -90.2 \\ %4.73  - 20110809
			& 55783.92105359 & T10 & 700  & 1.00  & 0.8952 & 88.3 & -71.6 \\ %5.48  - 20110810
			& 55783.93034186 & T10 & 700  & 1.00  & 0.9059 & 98.8 & -60.4 \\ %4.09  - 20110810
			& 55783.93904843 & T10 & 700  & 1.00  & 0.9159 & 77.5 & -50.4 \\ %3.09  - 20110810
			& 55784.91226546 & T10 & 700  & 1.01  & 0.0384 & -8.9 & 29.4 \\ %3.32  - 20110811
			& 55784.92096450 & T10 & 700  & 1.00  & 0.0484 & 6.7 & 40.1 \\ %4.67  - 20110811
			& 55784.92966123 & T10 & 700  & 1.00  & 0.0585 & -22.5 & 65.0 \\ %3.81  - 20110811
%			& 56087.96069884 & T10 & 1500  & 1.33  & 8.0  &  &  &  \\ %7.97  - 20120609
			& 56092.06873895 & T10 & 1100  & 1.01  & 0.1263 & -33.3 & 95.9 \\ %9.28  - 20120613
			& 56092.08211483 & T10 & 1100  & 1.00  & 0.1417 & -38.6 & 88.4 \\ %7.02  - 20120613
			& 56092.09549650 & T10 & 1100  & 1.00  & 0.1571 & -32.2 & 82.6 \\ %5.78  - 20120613
			& 56092.11057526 & T10 & 1200  & 1.00  & 0.1752 & -14.5 & 119.3 \\ %5.39  - 20120613
			& 56092.12515219 & T10 & 1200  & 1.01  & 0.1920 & -33.1 & 134.1 \\ %6.55  - 20120613
			& 56092.13970365 & T10 & 1200  & 1.03  & 0.2088 & -46.8  & 90.6\\ %5.88  - 20120613
			& 56115.08250551 & T07 & 1800  & 1.03  & -- & -- & -- \\ %8.85  - 20120706
\\  [-2ex]
19f-4-05194	& 55782.98843010 & T10 & 1200  & 1.03  & 0.6870 & 112.9 & -157.4 \\ %2.77  - 20110809
			& 55783.00298503 & T10 & 1200  & 1.06  & 0.7117 & 103.2 & -- \\ %2.66  - 20110809
			& 55783.01753302 & T10 & 1200  & 1.09  & 0.7364 & 74.6 & -158.1 \\ %4.05  - 20110809
			& 55783.87315996 & T10 & 1200  & 1.05  & 0.1874 & -112.9 & 133.3 \\ %4.53  - 20110810
			& 55783.89008688 & T10 & 1200  & 1.03  & 0.2161 & -119.2 & 130.2 \\ %2.84  - 20110810
			& 55783.90463660 & T10 & 1200  & 1.00  & 0.2408 & -105.2 & -- \\ %2.26  - 20110810
			& 56088.06386718 & T10 & 1000  & 1.02  & 0.1711 & -136.5 & 146.5 \\ %3.12  - 20120609
			& 56088.07611900 & T10 & 1000  & 1.01  & 0.1919 & -131.3 & 178.7 \\ %2.75  - 20120609
			& 56088.08878525 & T10 & 1000  & 1.00  & 0.2133 & -101.5 & 135.3 \\ %4.15  - 20120609
			& 56088.10219008 & T10 & 1000  & 1.00  & 0.2361 & -75.3 & 135.7 \\ %2.90  - 20120609
			& 56088.11444479 & T10 & 1000  & 1.00  & 0.2569 & -117.2 & 104.4 \\ %3.77  - 20120609
			& 56088.12670124 & T10 & 1000  & 1.01  & 0.2776 & -90.7 & 155.6 \\ %2.70  - 20120609
			& 56088.98096796 & T10 & 1100  & 1.21  & 0.7275 & 88.3 & -113.2 \\ %2.77  - 20120610
			& 56088.99448449 & T10 & 1100  & 1.16  & 0.7504 & 103.3 & -131.3 \\ %3.45  - 20120610
			& 56089.00793504 & T10 & 1100  & 1.12  & 0.7732 & 116.8 & -125.3 \\ %2.05  - 20120610
			& 56090.99015047 & T10 & 1000  & 1.15  & 0.1355 & -93.2 & 151.3 \\ %5.52  - 20120610
			& 56091.00237277 & T10 & 1000  & 1.12  & 0.1563 & -89.1 & 136.6 \\ %6.00  - 20120610
			& 56091.01459391 & T10 & 1000  & 1.08  & 0.1770 & -120.1 & 135.1 \\ %5.71  - 20120610
			& 56091.02789165 & T10 & 1000  & 1.06  & 0.1995 & -103.3 & 126.2 \\ %6.15  - 20120610
			& 56091.04011395 & T10 & 1000  & 1.04  & 0.2203 & -94.1 & 111.7 \\ %5.98  - 20120610
			& 56091.05233567 & T10 & 1000  & 1.02  & 0.2410 & -83.7 & 120.7 \\ %4.48  - 20120610
			& 56091.98183247 & T10 & 1100  & 1.17  & 0.8187 & 108.1 & -125.1 \\ %3.72  - 20120613
			& 56091.99521299 & T10 & 1100  & 1.13  & 0.8414 & 71.7 & -129.5 \\ %4.11  - 20120613
			& 56092.00859060 & T10 & 1100  & 1.09  & 0.8641 & 87.8 & -106.4 \\ %4.32  - 20120613
			& 56114.99812036 & T07 & 2100  & 1.01  & -- & -- & -- \\ %7.91  - 20120706
\\  [-2ex]
19g-2-08064	& 56087.98405758 & T10 & 1000  & 1.22  & 0.1398 & 74.8 & -92.1 \\ %5.03  - 20120609
			& 56087.99634992 & T10 & 1000  & 1.17  & 0.1469 & 56.4 & -101.0 \\ %5.63  - 20120609
			& 56088.00865904 & T10 & 1000  & 1.13  & 0.1541 & 56.2 & -105.7 \\ %8.29  - 20120609
			& 56088.02227224 & T10 & 1000  & 1.10  & 0.1620 & 74.1 & -82.5 \\ %8.51  - 20120609
            & 56088.03456978 & T10 & 1000  & 1.07  & 0.1691 & 46.0 & -106.4 \\ %5.03  - 20120609
			& 56088.04681292 & T10 & 1000  & 1.05  & 0.1762 & 72.6 & -97.9 \\ %6.52  - 20120609
			& 56088.93719029 & T10 & 1100  & 1.47  & 0.6940 & -92.9 & 88.9 \\ %8.29  - 20120610
			& 56088.95108885 & T10 & 1100  & 1.38  & 0.7021 & -98.3 & 91.4 \\ %5.74  - 20120610
			& 56088.96452203 & T10 & 1100  & 1.30  & 0.7099 & -101.2 & 94.1 \\ %6.10  - 20120610
			& 56089.06693907 & T10 & 1100  & 1.02  & 0.7694 & -110.7 & 79.8 \\ %7.19  - 20120610
			& 56089.08031379 & T10 & 1100  & 1.01  & 0.7772 & -102.6 & 90.0 \\ %9.44  - 20120610
			& 56089.09369662 & T10 & 1100  & 1.00  & 0.7850 & -100.7 & 88.6 \\ %9.63  - 20120610
			& 56115.13183924 & T07 & 1800  & 1.14  & -- & -- & -- \\ %12.30  - 20120706

\hline %\hline
\end{longtable}
\begin{list}{}{}
\item[]{\footnotesize{ {\bf Note.} The spectra observed with grating T10 have a mean signal-to-noise ratio (SNR), measured at the continuum, of 2.3 and 5.0 for the 17hr and 19hr fields, respectively.}}
\end{list}
\twocolumn
\end{center}

\subsection{Radial velocity measurements}

The intermediate-resolution spectroscopic data were used to measure radial velocities of each EB candidate. The presence of a double-lined H${\rm \alpha}$ in emission in the observed spectra allowed the measurement of RVs from both components. 

Firstly, RV corrections were computed with IRAF's RVCORRECT (RV package), where we set the Solar velocity (VSUN) to be $0$ km s$^{-1}$ in order to obtain heliocentric systemic velocities of the objects. 
We then used FXCOR (IRAF's RV package) to compute radial velocities via Fourier cross correlation. Due to the lack of other emission lines in our spectra, we applied the cross-correlation function (CCF) only around the H${\rm \alpha}$-line region. This was done to improve the CCF signal-to-noise ratio (SNR) since this feature appears in emission in all cases. For that, a single-Gaussian template covering the same wavelength region was generated to be used as cross-correlation kernel. This template is totally flat, except at the H${\rm \alpha}$ region where there is a Gaussian with full-width at half maximum (FWHM) given by the instrumental resolution, measured from arc lines images. This procedure aims at measuring the velocity shifts of narrow emission components in the observed H${\rm \alpha}$ profile.

%The obtained velocities for each stellar component of the investigated binaries are shown in tables \ref{obs17h} and \ref{obs19h}, as well as the orbital phase of the observations.
The obtained velocities for each stellar component of the investigated binaries are shown in columns 7 and 8 of table \ref{obs17-19h}, as well as the orbital phase of the observations, in column 6.

\section{Characterization of the low-mass eclipsing binaries}
\label{sec:charac}

The characterization of our LMEBs was performed in three different but complementary ways: a photometric characterization through a SED fitting, a spectral typing using spectral indices, and a spectroscopic characterization through a comparison to a library of synthetic spectra.

\subsection{Broad-band photometric characterization with SED fitting}\label{sect:sed}

\begin{table*}
\begin{center}
%\setcaptionmargin{1cm}
\caption{LMEBs coordinates and the broad-band photometry used to perform the SED fitting. The EBs from 19hr field have additional photometry from SDSS DR7, given in AB magnitudes. All other magnitudes are given in the Vega system.}

\label{tab_filters}
\begin{tabular}{lccccc}
\hline \hline \\ [-3ex]
Filter & 17e-3-02003 & 17h-4-01429 & 19c-3-08647 & 19f-4-05194 & 19g-2-08064 \\
%[0.5ex] 
\hline\\  [-2ex]

${\rm R.A.}$ {\scriptsize (J2000)} & 17:13:01.2 & 17:15:39.6 & 19:37:13.6 & 19:31:15.0 & 19:36:40.6 \\ 
${\rm Dec}$ {\scriptsize (J2000)} & $+$03:50:40.2 & $+$03:47:09.7 & $+$36:48:54.1 & $+$36:35:25.9 & $+$36:09:45.8 \\ 
\\  [-2ex]
SDSS-u   & -- & -- & 21.026$\pm$0.094 & 21.60$\pm$0.18 & 20.057$\pm$0.051 \\ 
SDSS-g   & -- & -- & 18.623$\pm$0.008 & 19.188$\pm$0.011 & 17.729$\pm$0.005 \\ 
SDSS-r   & -- & -- & 17.339$\pm$0.005 & 17.983$\pm$0.007 & 16.520$\pm$0.004 \\ 
SDSS-i   & -- & -- & 16.537$\pm$0.004 & 17.436$\pm$0.006 & 15.962$\pm$0.004 \\ 
SDSS-z   & -- & -- & 16.083$\pm$0.007 & 17.107$\pm$0.012 & 15.592$\pm$0.006 \\ 
UKIDSS-Z & 16.114$\pm$0.049 & 16.93$\pm$0.078 & 15.609$\pm$0.035 & 16.802$\pm$0.096 & 15.275$\pm$0.018 \\ 
UKIDSS-Y & 15.808$\pm$0.097 & 16.48$\pm$0.18 & 15.321$\pm$0.054 & 16.67$\pm$0.17 & 14.993$\pm$0.034 \\
UKIDSS-J & 15.272$\pm$0.060 & 15.84$\pm$0.12 & 14.812$\pm$0.030 & 16.013$\pm$0.091 & 14.466$\pm$0.017 \\
UKIDSS-H & 14.718$\pm$0.017 & 15.148$\pm$0.019 & 14.210$\pm$0.002 & 15.320$\pm$0.006 & 13.858$\pm$0.002 \\
UKIDSS-K & 14.407$\pm$0.027 & 14.904$\pm$0.034 & 14.001$\pm$0.010 & 15.167$\pm$0.032 & 13.673$\pm$0.002 \\
2MASS-J  & 15.315$\pm$0.059 & 15.888$\pm$0.075 & 14.877$^{\rm *}$ & 16.053$\pm$0.075 & 14.477$\pm$0.029 \\
2MASS-H  & 14.575$\pm$0.068 & 15.001$\pm$0.094 & 14.258$\pm$0.054 & 15.272$\pm$0.076 & 13.863$\pm$0.033 \\
2MASS-Ks & 14.434$\pm$0.078 & 14.93$\pm$0.15 & 13.918$^{\rm *}$ & 15.23$\pm$0.17 & 13.715$\pm$0.050 \\
WISE-W1  & 14.215$\pm$0.028 & 14.777$\pm$0.033 & 13.730$\pm$0.030 & 15.061$\pm$0.043 & 13.608$\pm$0.027 \\
WISE-W2  & 14.172$\pm$0.048 & 14.769$\pm$0.078 & 13.678$\pm$0.040 & 15.76$\pm$0.22 & 13.686$\pm$0.044 \\
%WISE-W3  & 12.3610$\pm$0.39000 & 12.2910 & 12.5770$\pm$0.349000 & 12.9310 & 12.8530 \\
%WISE-W4  & 9.28500 & 9.21800 & 9.22700 & 9.07700 & 9.45200 \\

\hline %\hline
\end{tabular}
\begin{list}{}{}
\item[]{\scriptsize{{\bf Note.} $^{\rm *}$ These values are only upper limits.}}
\end{list}
\end{center}
\end{table*}

\begin{table*}
\begin{center}
%\setcaptionmargin{1cm}
\caption{Atmospheric parameters and spectral types for our 5 LMEBs obtained from three different methods: SED fitting (columns 2-3), spectral indices (columns 4-7), and comparison to synthetic spectra (columns 8-12). The errors are of $100$ K for $T_{\rm eff}$ and $0.5$ for $\log g$. The spectral types have an uncertainty of one subclass.}
\label{tab_charac}
\begin{tabular}{cccccccccccc}
%    \toprule
\hline \hline \\ [-3ex]
 & \multicolumn{2}{c}{SED fitting} & \multicolumn{4}{c}{Spectral indices} & \multicolumn{5}{c}{Synthetic spectra} \\
\cmidrule(lr){2-3} \cmidrule(lr){4-7} \cmidrule(lr){8-12}
object & $T_{\rm eff}$ (K) & $\log  g$ & {Ratio A} & {TiO-a} & {$\Re$-index} & type$^{ \rm a}$ & $T_{\rm eff,1}$ (K) & $\log  g_{\rm 1}$ & $T_{\rm eff,2}$ (K) & $\log  g_{\rm 2}$ & type$^{ \rm b}$ \\
%[0.5ex] 
\hline\\  [-2ex]
%17e-3-02003	& 3600 & 5.5 & M1 & M1 & M1 & 3600 & 5.5 & 3600 & 5.5 & M1+M1 \\
17e-3-02003	& 3600 & 5.5 & 1.108 & 1.111 & 1.267 & M1 & 3800 & 4.5 & 3500 & 4.5 & M0+M2.5 \\
%17h-4-01429	& 3500 & 4.0 & M3 & M3.5 & M3.5 & 3500 & 4.0 & 3500 & 4.0 & M3+M4 \\
17h-4-01429	& 3500 & 4.0 & 1.228 & 1.509 & 2.133 & M3.5 & 3400 & 4.5 & 3200 & 5.0 & M3+M4 \\
%19c-3-08647	& 3800 & 5.0 & M1 & M1.5 & M1 & 3800 & 5.0 & 3800 & 5.0 & M0+M3 \\
19c-3-08647	& 3800 & 5.0 & 1.125 & 1.184 & 1.300 & M1 & 3900 & 4.5 & 3000 & 4.5 & M0+M5 \\
%19f-4-05194	& 4200 & 5.0 & K7 & K7 & K7 & 4200 & 5.0 & 4200 & 5.0 & K7+M1 \\
19f-4-05194	& 4200 & 5.0 & 1.027 & 1.034 & 1.090 & K7 & 4400 & 4.5 & 3500 & 5.0 & K5+M2.5 \\
%19g-2-08064 & 4200 & 5.0 & K7 & K7 & K7 & 4200 & 5.0 & 4200 & 5.0 & K7+M1 \\
19g-2-08064 & 4200 & 5.0 & 1.044 & 1.072 & 1.124 & K7 & 4200 & 4.5 & 3100$^{ \rm *}$ & 5.0 & K6+M4.5$^{ \rm *}$ \\

\hline %\hline
\end{tabular}
\begin{list}{}{}
\item[]{\scriptsize{{\bf Notes.} $^{\rm a}$ EB's spectral type from measured spectral indices. $^{\rm b}$ EB spectral-type combination according to $T_{\rm eff}$.} $^{\rm *}$ This value is discrepant with respect to the expected effective temperature and should be taken with caution (for more details, see sect. \ref{sec:res}).}
\end{list}
\end{center}
\end{table*}

\begin{figure*}
\begin{center}
%\setcaptionmargin{1cm}
\includegraphics[width=0.8\columnwidth,angle=90]{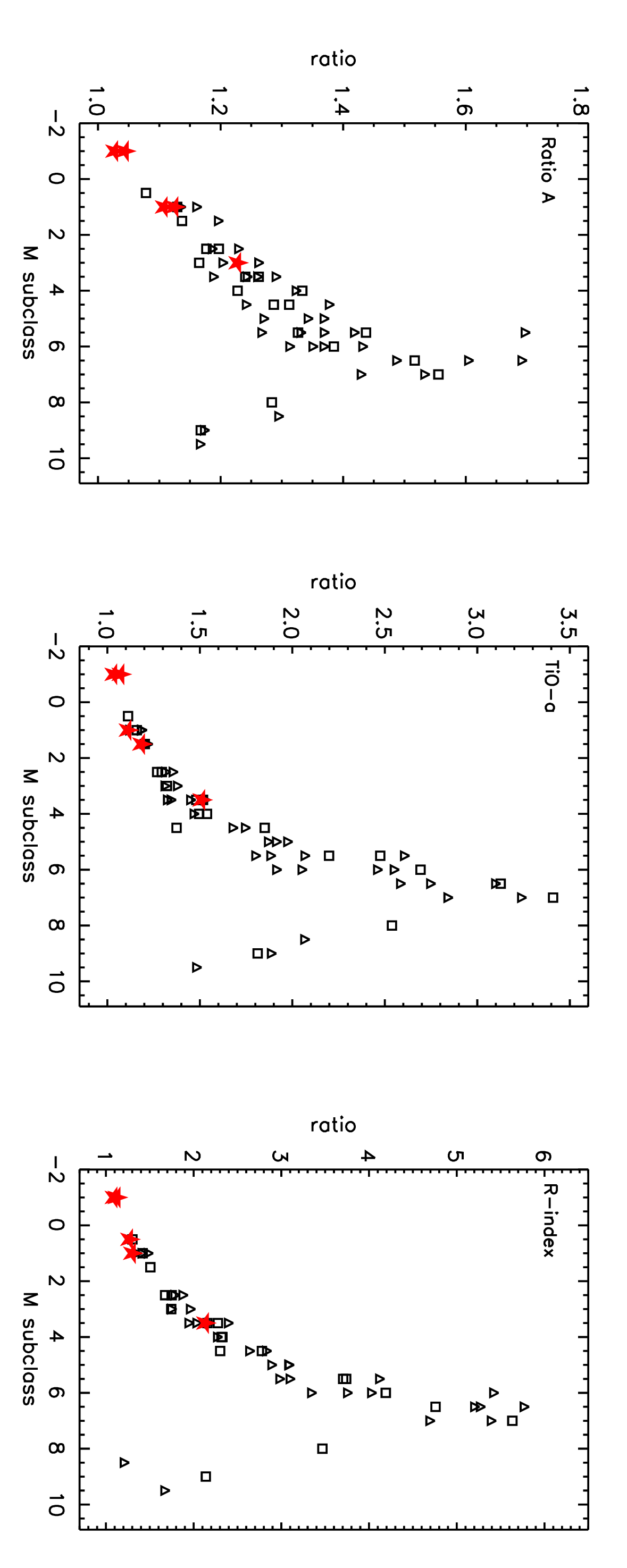}
\caption{Stellar spectral types versus spectral indices. The M-dwarf templates from \citet{Leggett00} are represented by open triangles and the templates from \citet{Cruz02} are represented by open squares. The filled stars are the 5 LMEBs from this work. The left panel shows the {\it Ratio A} index \citep{Kirkpatrick91}, the central panel shows the {\it TiO-a} index \citep{Kirkpatrick99}, and the right panel shows the {\it $\Re$-index} \citep{Aberasturi14}.}
\label{figSpecInd}
\end{center}
\end{figure*}

In order to confirm at first the low-mass nature of the binaries in study, we used the Virtual Observatory tool called VOSA\footnote{VOSA is an open VO tool that can be accessed at http://svo2.cab.inta-csic.es/theory/vosa/} \citep[Virtual Observatory SED Analyzer,][]{Bayo08} to perform a spectral energy distribution fitting. For that, we used the broad-band photometry available in the literature as well as those obtained from the WFCAM photometry. For all selected 5 LMEBs, we considered the broadband photometry from WFCAM Z, Y, J, H, K \citep[Wide-Field Camera, ][]{Hodgkin09}, 2MASS J, H, K \citep[Two Micron All Sky Survey, ][]{Skrutskie06}, and WISE W1, W2 \citep[Wide-field Infrared Survey Explorer, ][]{Wright10}. Additionally, for the 19hr field objects only, we also considered the available SDSS u, g, r, i, z filters \citep[Sloan Digital Sky Survey Data Release 7, ][]{York00}. The photometry data used are shown in table \ref{tab_filters}.

We then obtained an estimate for the effective temperature ($T_{\rm eff}$) of each binary system by using NextGen models \citep[][and references therein]{Baraffe98,Hauschildt99}, where the best fitting model was chosen through a $\chi^2$ minimization \citep[for details, see][]{Bayo08}. Note that we fixed the metallicity to be solar. The surface gravity ($\log g$) was also free to vary, although the SED fitting has a low sensitivity to this parameter and the values obtained from the fit should be taken with caution.

The obtained atmospheric parameters are presented in table \ref{tab_charac} (columns 2 and 3), with uncertainties corresponding to the step-size in the used model grid, which are $100$ K and $0.5$ for $T_{\rm eff}$ and $\log g$, respectively. These parameters were considered as first estimates and are not the final values adopted for each component of the binary. However, they are good approximations that confirm the expected low-mass nature of our objects.

\subsection{Spectroscopic characterization with spectral indices}

The acquired low-resolution spectra (described in sect. \ref{sec:specobs}) were used for spectral typing. As mentioned previously, the observed spectral range was chosen to include some features that are characteristic of the late-type stars, specially M dwarfs. These features, as for instance the titanium oxide (TiO) and vanadium oxide (VO) bands, are good markers to distinguish early- and late-M dwarfs \citep{Keenan52,Kirkpatrick91}.

Three spectral indices were selected to derive the spectral type of our LMEBs. The first index used is the {\it Ratio A} index (7020-7050 \AA / 6960-6990 \AA), defined by \citet{Kirkpatrick91}. This index is based on the calcium monohydride (CaH) molecule, which is a strong feature that appears in early-M stars and is a good luminosity indicator. 
The {\it TiO-a} index \citep[7033-7048 \AA / 7058-7073 \AA,][]{Kirkpatrick99} is the second one used, which is based on the TiO molecular band and is stronger for early- and mid-M dwarfs. Finally, the third index used is the {\it $\Re$-index} (7485-7515 \AA / 7120-7150 \AA), derived more recently by \citet{Aberasturi14}, and measures the minimum of the same TiO band.

As a comparative way to derive the spectral type of our targets, we have applied the same indices to a set of M-dwarf template spectra from the literature. A total of 58 M-dwarf templates, 39 spectra from \citet{Leggett00} and 19 spectra from \citet{Cruz02} were used, covering M-dwarf subclasses from M0.5 V to M9.5 V. All spectra, templates and those from this work, were normalized to a pseudo-continuum before measuring all indices to avoid any flux calibration discrepancy. The measured values for our 5 LMEBs are presented in table \ref{tab_charac} (columns 4 to 6). In column 7, we show the assigned spectral type for the binary, which is the averaged type obtained from all indices, with an uncertainty of one subclass. It is worth to mention that the spectral type defined here is not assigned to each component of the binary but to the composite spectrum.

Figure \ref{figSpecInd} shows the behavior of the spectral types versus the measured values for the three used indices. The open triangles represent the templates from \citet{Leggett00} and the open squares are templates from \citet{Cruz02}. Our objects are illustrated as filled stars. The y-axis displays the value measured for each index ({\it Ratio A} on the left panel, {\it TiO-a} at the center, and {\it $\Re$-index} on the right panel). The x-axis shows all M subclasses. Note that the numerical scale here represents each M-dwarf subclass, going to M0 to M9, where $0$ is the M0 subclass, $1$ is the M1 subclass, successively. Negative numbers represent earlier types, where $-1$ represents the K7 subclass and $-2$ represents the K5 subclass. This convention was used by \citet[][fig. 6]{Kirkpatrick91} and it was also adopted here.

\subsection{Spectroscopic characterization by comparison to synthetic spectra}
\label{lowres}

The last procedure for the characterization of our objects was to perform a comparison to a library of synthetic spectra, using our observed flux-calibrated low-resolution spectra. For that, we selected the BT-Settl model spectra from \citet{Allard12,Allard13}\footnote{All synthetic spectra used are available at https://phoenix.ens-lyon.fr/Grids/BT-Settl/}, computed with the updated solar abundances from \citet{Asplund09}. This library uses the PHOENIX code for 1D-model atmospheres \citep{Allard94,Allard97,Allard01} to generate high-resolution spectra ($0.02$ \AA, at optical wavelengths). The BT-Settl models were chosen because they are valid for all temperatures, reaching therefore the low-temperature regime of our objects.

We have composed a grid of BT-Settl spectra with effective temperatures ranging from $3000$ to $5000$ K, with a step of $100$ K, always assuming solar metallicity (${\rm [Fe/H]=0.0}$) and no ${\rm \alpha}$-element enhancement (${\rm [\alpha/H]=0.0}$). The surface gravities were 4.5 and 5.0. Before performing the comparison, the whole grid was degraded to the same resolution as our spectroscopic data.

\begin{figure}
\begin{center}
%\setcaptionmargin{1cm}
\includegraphics[width=1.0 \columnwidth]{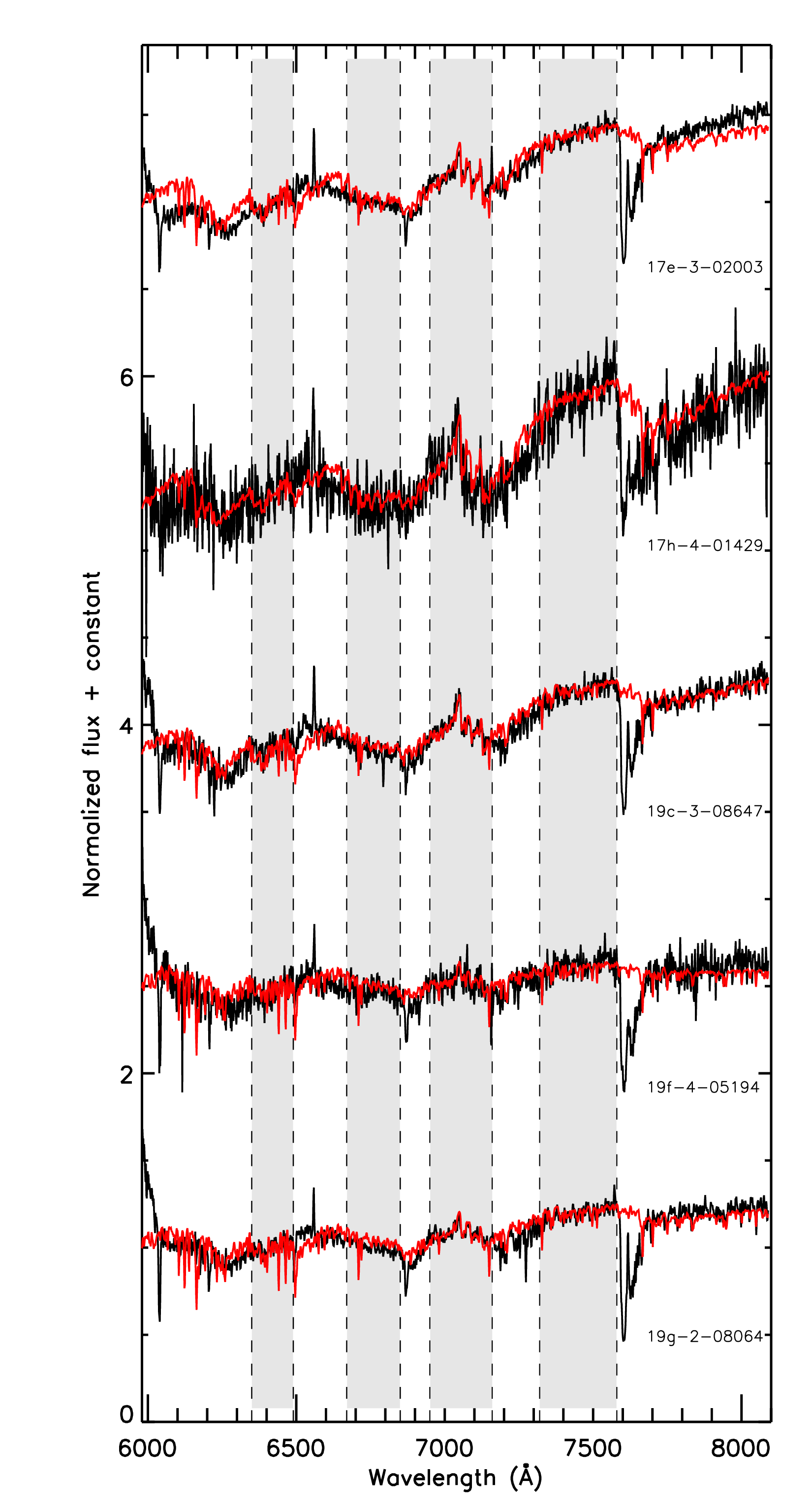}
\caption{Observed low-resolution spectra (black) and the best-fitting model (red) obtained from a combination of two synthetic spectra. All spectra are flux-calibrated and normalized to 1 at $\lambda = 7500$ \AA. The atmospheric parameters of these models are presented in table \ref{tab_charac} (columns 8 to 11). The gray regions show the wavelength intervals used for the comparison.}
\label{figSynthSpec}
\end{center}
\end{figure}

With the objective of determining the effective temperatures of both components of the system, two synthetic spectra were combined to reproduce the observed spectrum of the binary in question. It is worth to mention that, within the observed spectral range, there are some telluric absorption features that should not be accounted for the fit, such as telluric ${\rm H{_2}O}$ and ${\rm O{_2}}$ molecular bands. Hence, we have performed the comparison considering only a few windows to exclude such telluric contamination. We also took into account that the borders of our observed spectra present flux-calibration issues, thus, they were also excluded from the fit. This way, we applied our comparison method only for the following wavelength intervals: from $6350$ to $6490$ \AA, from $6670$ to $6850$ \AA, from $6950$ to $7160$ \AA, and from $7320$ to $7580$ \AA. Note that the region around the H${\rm \alpha}$ line was also excluded since this feature appears in emission in all observed spectra. It is also important to have in mind that a synthetic spectrum can not perfectly reproduce an observed spectrum, therefore, some discrepancies are still present within the chosen wavelength intervals.

The best-fitting model was obtained via $\chi^2$ minimization, providing a pair of parameters ($T_{\rm eff}$, $\log g$) for each component of the binary. In this procedure, the $T_{\rm eff}$ was set free to vary within the range specified earlier. However, the $\log g$ values for the primary and secondary components were fixed to the values obtained from the light-curve modeling\footnote{The used $\log g$ values were obtained from previously performed trial runs of the light-curve modeling code.}. The light-curve modeling procedure is described later on Sect. \ref{solve}. We adopted nearest 0.5 dex values for $\log g$, for instance, for a star with a derived surface gravity of $4.8$ we used a synthetic spectra computed with $\log g=5.0$.

The derived atmospheric parameters for all LMEB systems are shown in table \ref{tab_charac} (columns 8 to 11), with errors of $100$ K and $0.5$ for the effective temperature and the surface gravity, respectively (the step-size in the model grid). 
In column 12, we present the composition of spectral types for each binary system. These spectral types were determined on the basis of the resulting temperatures and the effective temperature scale published by \citet{Pecaut13}. 
It is important to mention that the $T_{\rm eff}$ value obtained for the secondary component of EB 19g-2-08064 is highly discrepant regarding the value expected from the phase-folded light curve, which will be discussed later on sect. \ref{sec:res}. For this reason, this particular temperature should be taken with caution.

To illustrate our results, figure \ref{figSynthSpec} presents the best-fitting model for the 5 LMEB systems from the combination of two synthetic spectra. All spectra are flux-calibrated and normalized to 1 at $\lambda = 7500$ \AA, where an arbitrary constant was added for visualization. The observed spectra are shown in black and the models are presented in red. The regions featured in gray show the wavelength intervals sampled to perform the fitting.

\section{Solving the 5 new LMEB systems}\label{solve}

\subsection{Modeling light curves and radial velocity shifts}

All five LMEB systems were solved through the analysis of the photometric and spectroscopic data combined. 
The light curves and the measured radial velocity shifts were simultaneously modeled with the JKTEBOP\footnote{JKTEBOP is a modified version of EBOP \citep[Eclipsing Binary Orbit Program,][]{Etzel81,Popper81}, available at http://www.astro.keele.ac.uk/jkt/codes/jktebop.html.} code \citep{Southworth04,Southworth13}. This code was considered as the best option for the analysis of our systems, since it provides integrated fitting of light-curves with radial velocity data and is suitable for nearly spherical stars in detached binaries.

\begin{figure*}
\begin{center}
%\setcaptionmargin{1cm}
\includegraphics[width=0.75 \columnwidth, angle=90]{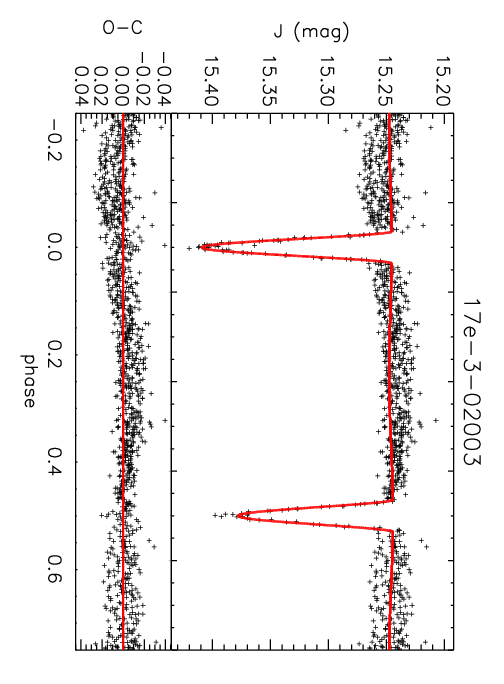}
\includegraphics[width=0.75 \columnwidth, angle=90]{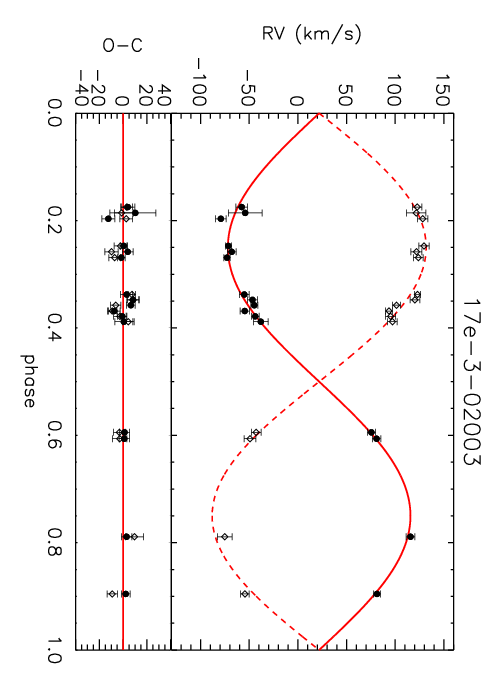}
\includegraphics[width=0.75 \columnwidth, angle=90]{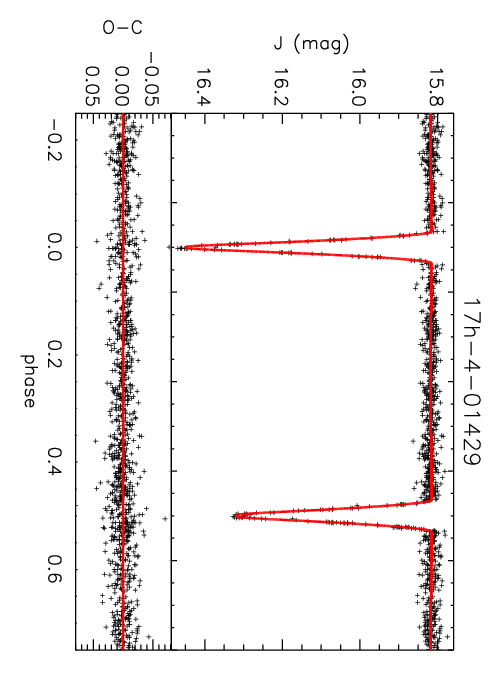}
\includegraphics[width=0.75 \columnwidth, angle=90]{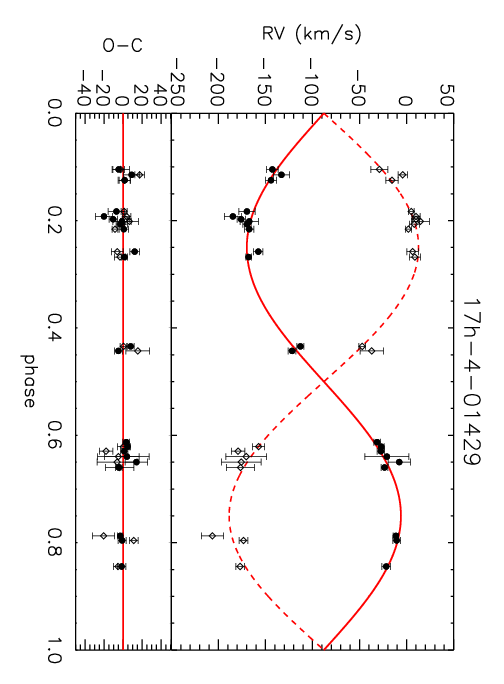}
\caption{Light curves (left) and radial velocity shifts (right) of the LMEBs from the 17hr field as function of the orbital phase. The observational data are presented in black and the best-fitting models from the JKTEBOP code are shown in red. The obtained parameters are listed in table \ref{tab_res}. Each individual plot brings below itself the residuals from the fit (O-C).}
\label{figmodelfit17hr}
\end{center}
\end{figure*}

\begin{figure*}
\begin{center}
%\setcaptionmargin{1cm}
\includegraphics[width=0.75 \columnwidth, angle=90]{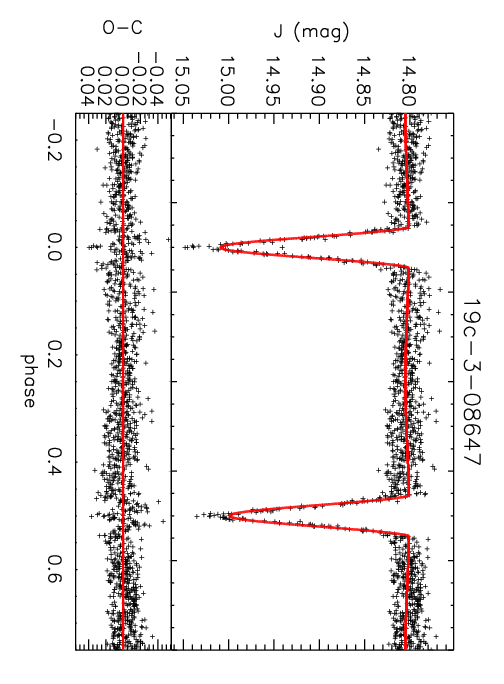}
\includegraphics[width=0.75 \columnwidth, angle=90]{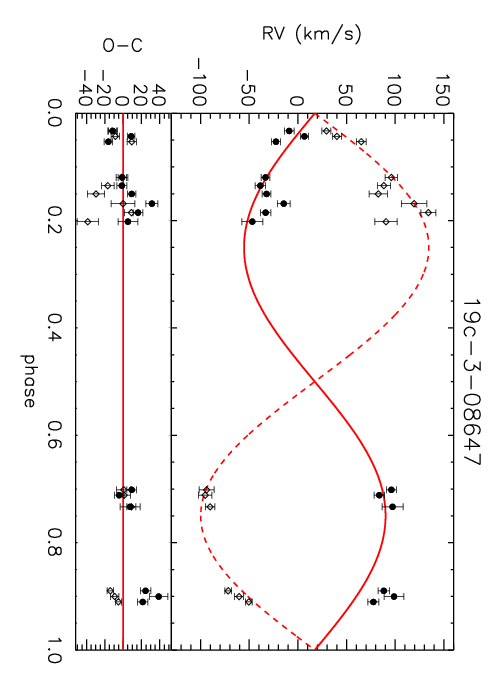}
\includegraphics[width=0.75 \columnwidth, angle=90]{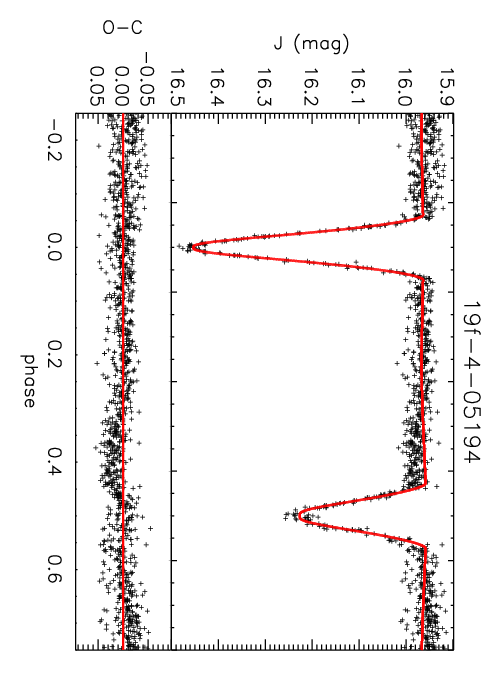}
\includegraphics[width=0.75 \columnwidth, angle=90]{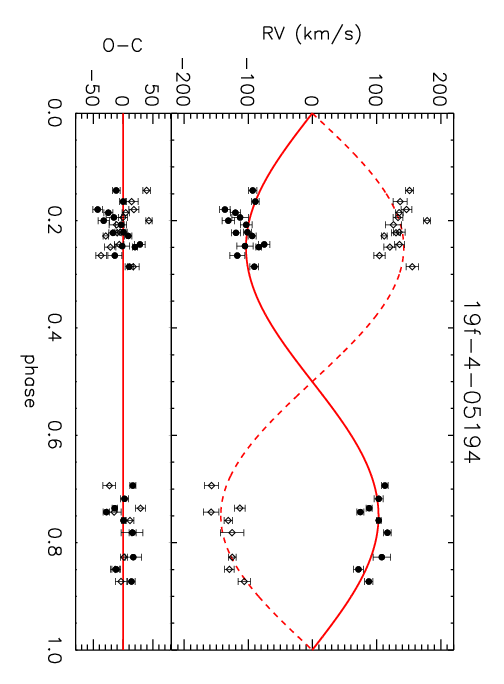}
\includegraphics[width=0.75 \columnwidth, angle=90]{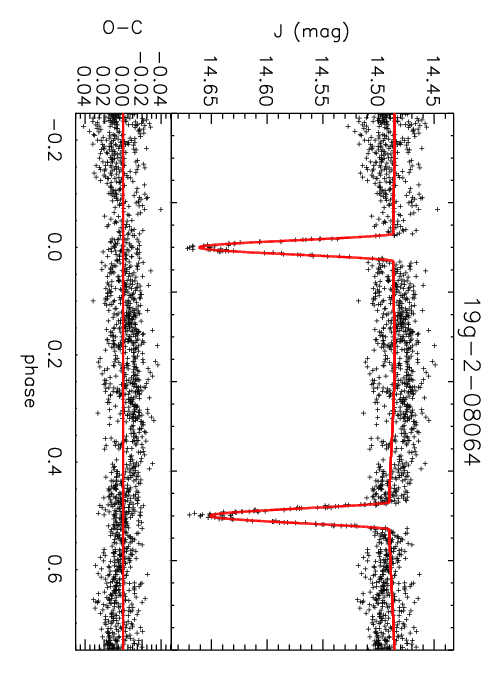}
\includegraphics[width=0.75 \columnwidth, angle=90]{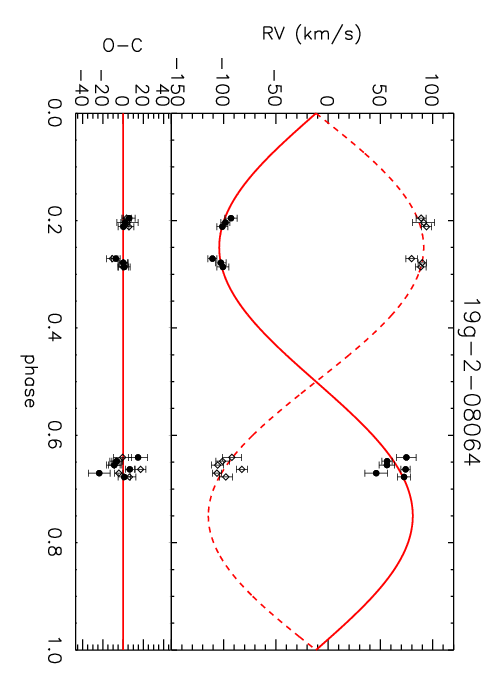}
\caption{Light curve (left) and radial velocity shifts (right) of the LMEBs from the 19hr field as function of the orbital phase. The details are the same as described in fig. \ref{figmodelfit17hr}.}
\label{figmodelfit19hr}
\end{center}
\end{figure*}

JKTEBOP uses the Nelson-Davis-Etzel model for eclipsing binaries \citep[NDE model,][]{Nelson72,Popper81} to model light curves of and the best-fitting model is chosen by a Levenberg-Marquardt minimization algorithm \citep[MRQMIN,][]{Press92}. 
Several parameters of the system were allowed to vary:
\begin{itemize}
\renewcommand\labelitemi{--}
\item {\em surface brightness ratio} ($J$),
\item {\em both stellar radii} ($r_{\rm 1}$, $r_{\rm 2}$), defined as $r_{\rm 1} = R_{\rm 1} / a$ and $r_{\rm 2} = R_{\rm 2} / a$, where $R$ is the stellar radius, $r$ is the fractional radius, and $a$ is the orbital separation,
\item {\em reference time of primary minimum} ($T_0$), given in days,
\item {\em orbital period} ($P_{\rm orb}$), also given in days,
\item {\em orbital inclination angle} ($i$), given in degrees, and
\item {\em light scale factor}, which defines the baseline level of the light curve, and in this case is given in magnitudes.
\end{itemize}
In fact, in order to derive the radii of both components ($r_{\rm 1}$, $r_{\rm 2}$), the code actually fits the radii sum and ratio, $r_{\rm 1}+r_{\rm 2}$ and $k$, respectively.

Some other light curve parameters were kept fixed to a certain value. The {\em photometric mass ratio} ($q_{\rm phot}$), which accounts for the small deformation of the objects, was determined from the expected mass of the stars based on a few trial runs in the code, and kept fixed. It is important to emphasize that the derived stellar masses result from the radial velocity curve analysis.  
We also used a fixed value for the {\em gravity-darkening coefficients} ($\beta$) of $0.32$, which is a reasonable value for late-type stars with convective envelopes \citep{Lucy67}, and since our light curves do not have enough precision to fit for this parameter.

We have adopted a square-root law for the limb darkening (LD) since, according to \citet{vanHamme93}, this law is more appropriate for infrared light curves. We used JKTLD\footnote{http://www.astro.keele.ac.uk/jkt/codes/jktld.html.} code \citep{Southworth07} associated with JKTEBOP, to obtain limb-darkening coefficients by interpolating the values within theoretical tables. We selected the values computed by \citet{Claret00}, which are based on PHOENIX models \citep{Allard97} and are more suitable for our temperature range. 
For that aim, we considered the atmospheric parameters ($T_{\rm eff}$, $\log g$) obtained from the analysis of low-resolution spectra (see Sect. \ref{lowres}). We also considered a solar metallicity and a microturbulence velocity of $2.0$ km s$^{-1}$. These coefficients were also kept fixed due to the limited precision of the photometric data.

A few assumptions were made when performing the model fit, for instance, we assumed null orbital eccentricity for the systems. This is a good approximation, since the orbital periods of all LMEBs are small enough to consider that the systems are circularized by tidal forces. We also considered the reflection coefficients to be initially zero, however, we allowed JKTEBOP to adjust these coefficients depending on the system geometry. Finally, we also assumed the absence of a third light source in the system.

The latest version of the JKTEBOP code \citep{Southworth13}, used for the present analysis, also models RV measurements simultaneously, where the following paramaters were fitted:
\begin{itemize}
\renewcommand\labelitemi{--}
\item {\em both velocity amplitudes} ($K_{\rm 1}$, $K_{\rm 2}$), and
\item {\em the systemic velocity} ($\gamma_{\rm sys}$).
\end{itemize}
All obtained velocities are given in km s$^{-1}$. It is important to emphasize that $\gamma_{\rm sys}$ is a free parameter for the primary star only. In the case of the secondary component, the systemic velocity was kept fixed to the value found for the primary.

\subsection{Results}\label{sec:res}

The JKTEBOP code allows the user to choose between several internal algorithms designed for different purposes. Two of them were selected to model fitting and access the parameter uncertainties. First, a MRQMIN algorithm \citep{Press92} was performed  to find the best-fitting model via a least-squares optimization (JKTEBOP {\it task $3$}) for each binary. 

Figures \ref{figmodelfit17hr} and \ref{figmodelfit19hr} present the best-fitting models obtained for the LMEBs from the 17hr and 19hr fields, respectively. The WTS light curves are shown on the left as function of the orbital phase, where each cross (black) represents an individual observation. The radial velocity curves are illustrated on the right, where the filled circles are the RV measurements of the primary component and the open diamonds are RVs of the secondary component. The red solid and dashed lines show the best-fitting RV models obtained with JKTEBOP. The residuals from the fit (O-C) are presented at the lower panel of each plot.

%\begin{landscape}
\begin{table*}
\begin{center}
\caption{Summary of the results obtained for the 5 LMEBs.% The errors obtained for the periods are shown in brackets, where the number represents the uncertainty of the last decimal place in the given period.
}
\label{tab_res}
\begin{tabular}{lccccc}
%    \toprule
\hline \hline \\ [-3ex]
Binary  & 17e-3-02003 & 17h-4-01429 & 19c-3-08647 & 19f-4-05194 & 19g-2-08064 \\
%[0.5ex] 
\hline\\  [-2ex]

\multicolumn{6}{l}{\scriptsize FITTED PARAMETERS}  \\

${\rm J}$ & $0.783 \pm 0.028$ & $0.8677 \pm 0.0051$ & $0.898 \pm 0.027$ & $0.5660 \pm 0.0044$ & $0.8492 \pm 0.0111$ \\
$(R_{\rm 1}+R_{\rm 2})/{a}$ & $0.2304 \pm 0.0048$ & $0.17931 \pm 0.00084$ & $0.2783 \pm 0.0038$ & $0.3745 \pm 0.0019$ & $0.1955 \pm 0.0021$ \\
${k}$ & $0.88 \pm 0.35$ & $0.8189 \pm 0.0081$ & $0.86 \pm 0.30$ & $0.6531 \pm 0.0048$ & $0.95 \pm 0.24$ \\
$R_{\rm 1}/a$ & $0.123 \pm 0.019$ & $0.09858 \pm 0.00051$ & $0.150 \pm 0.021$ & $0.2266 \pm 0.0010$ & $0.1002 \pm 0.0117$ \\
$R_{\rm 2}/a$ & $0.108 \pm 0.023$ & $0.08073 \pm 0.00070$ & $0.128 \pm 0.023$ & $0.1480 \pm 0.0013$ & $0.0954 \pm 0.0134$ \\
${i}$ {($^\circ $)} & $81.77 \pm 0.43$ & $89.15 \pm 0.13$ & $81.29 \pm 0.39$ & $85.62 \pm 0.20$ & $83.33 \pm 0.20$ \\
{Light scale} {($mag$)} & $15.2503 \pm 0.0006$ & $15.8191 \pm 0.0003$ & $14.8092 \pm 0.0006$ & $15.9724 \pm 0.0004$ & $14.4830 \pm 0.0005$ \\
${T_{\rm 0}}$ {\scriptsize (MHJD$-54000.$)} {($d$)} & $553.12520 \pm 0.00027$ & $553.15187 \pm 0.00012$ & $318.00600 \pm 0.00010$ & $317.60678 \pm 0.00014$ & $318.30802 \pm 0.00015$ \\
%${P_{\rm orb}}$ {($d$)} & $1.2250078 \pm 0.0000003$ & $1.4445891 \pm 0.0000002$ & $0.86746584 \pm 0.00000008$ & $0.58953012 \pm 0.00000008$ & $1.7204091 \pm 0.0000003$ \\
%${P_{\rm orb}}$ {($d$)} & $1.2250078(3)$ & $1.4445891(2)$ & $0.86746584(8)$ & $0.58953012(8)$ & $1.7204092(3)$ \\
${P_{\rm orb}}$ {($d$)} & $1.2250078$ $\pm$ & $1.4445891$ $\pm$ & $0.86746584$ $\pm$ & $0.58953012$ $\pm$ & $1.7204091$ $\pm$ \\
              & $0.0000003$ & $0.0000002$ & $0.00000008$ & $0.00000008$ & $0.0000003$ \\

${K_{\rm 1}}$ {($km$ $s^{-1}$)} & $93.80 \pm 1.45$ & $81.89 \pm 1.15$ & $72.70 \pm 2.07$ & $103.28 \pm 1.40$ & $92.56 \pm 1.73$ \\
${K_{\rm 2}}$ {($km$ $s^{-1}$)} & $109.93 \pm 1.71$ & $100.71 \pm 1.43$ & $117.19 \pm 2.22$ & $142.61 \pm 1.81$ & $102.99 \pm 1.71$ \\
${\gamma_{\rm sys}}$ {($km$ $s^{-1}$)} & $21.86 \pm 1.01$ & $-87.89 \pm 0.79$ & $17.37 \pm 1.03$ & $-0.33 \pm 1.07$ & $-11.63 \pm 1.15$ \\

\\ [-2ex]
\multicolumn{6}{l}{\scriptsize ABSOLUTE DIMENSIONS}  \\

${L_{\rm 2}/L_{\rm 1}}$ & $0.62 \pm 0.60$ & $0.584 \pm 0.011$ & $0.69 \pm 0.58$ & $0.264 \pm 0.004$ & $0.84 \pm 0.43$ \\
${a}$ {($R_{\odot}$)} & $4.982 \pm 0.049$ & $5.212 \pm 0.053$ & $3.292 \pm 0.054$ & $2.872 \pm 0.027$ & $6.692 \pm 0.082$ \\
${q}$ & $0.853 \pm 0.020$ & $0.813 \pm 0.016$ & $0.620 \pm 0.021$ & $0.724 \pm 0.013$ & $0.899 \pm 0.023$ \\
${M_{\rm 1}}$ {($M_{\odot}$)} & $0.597 \pm 0.020$ & $0.503 \pm 0.016$ & $0.393 \pm 0.019$ & $0.531 \pm 0.016$ & $0.717 \pm 0.027$ \\
${M_{\rm 2}}$ {($M_{\odot}$)} & $0.510 \pm 0.016$ & $0.409 \pm 0.013$ & $0.244 \pm 0.014$ & $0.385 \pm 0.011$ & $0.644 \pm 0.025$ \\
${R_{\rm 1}}$ {($R_{\odot}$)} & $0.611 \pm 0.095$ & $0.514 \pm 0.006$ & $0.494 \pm 0.069$ & $0.651 \pm 0.007$ & $0.670 \pm 0.078$ \\
${R_{\rm 2}}$ {($R_{\odot}$)} & $0.54 \pm 0.11$ & $0.421 \pm 0.006$ & $0.422 \pm 0.077$ & $0.425 \pm 0.006$ & $0.638 \pm 0.090$ \\
${\log g_{\rm 1}}$ & $4.64 \pm 0.14$ & $4.717 \pm 0.008$ & $4.65 \pm 0.13$ & $4.536 \pm 0.007$ & $4.64 \pm 0.10$ \\
${\log g_{\rm 2}}$ & $4.69 \pm 0.18$ & $4.801 \pm 0.010$ & $4.57 \pm 0.16$ & $4.766 \pm 0.009$ & $4.64 \pm 0.13$ \\
${\rho_{\rm 1}}$ {($g$ $cm^{-3}$)} & $2.6 \pm 1.6$ & $3.706 \pm 0.067$ & $3.3 \pm 1.8$ & $1.928 \pm 0.030$ & $2.38 \pm 0.83$ \\
${\rho_{\rm 2}}$ {($g$ $cm^{-3}$)} & $3.3 \pm 2.2$ & $5.49 \pm 0.16$ & $3.2 \pm 1.8$ & $5.01 \pm 0.14$ & $2.5 \pm 1.3$ \\

${\log L_{\rm 1}}$ {($L_{\odot}$)} & $-1.15 \pm 0.14$ & $-1.497 \pm 0.052$ & $-1.29 \pm 0.13$ & $-0.844 \pm 0.040$ & $-0.90 \pm 0.11$ \\
${\log L_{\rm 2}}$ {($L_{\odot}$)} & $-1.41 \pm 0.19$ & $-1.776 \pm 0.056$ & $-1.88 \pm 0.17$ & $-1.611 \pm 0.051$ & $-1.47 \pm 0.21$ \\
${M_{\rm bol,1}}$ {($mag$)} & $7.63 \pm 0.36$ & $8.49 \pm 0.13$ & $7.98 \pm 0.32$ & $6.86 \pm 0.10$ & $7.00 \pm 0.28$ \\
${M_{\rm bol,2}}$ {($mag$)} & $8.27 \pm 0.48$ & $9.19 \pm 0.14$ & $9.46 \pm 0.42$ & $8.78 \pm 0.13$ & $8.42 \pm 0.52$ \\
${d}$ {($pc$)} & $875 \pm 117$ & $805 \pm 62$ & $571 \pm 66$ & $1400 \pm 118$ & $731 \pm 75$ \\

\hline %\hline
\end{tabular}
\end{center}
\end{table*}
%\end{landscape}

Some of the phase-folded light curves present out-of-eclipse sinusoidal-like modulations, that could be due to ellipsoidal variations. We have then verified the sphericity of each binary component considering the Roche geometry. We have calculated the Roche equipotentials for each binary and we have found that for 4 EBs we do not expect any deformation (in a $10^{-8}$ level). Just one system, 19f-4-05194, presents a small deformation of around $0.2\%$ of the derived radius, which can be expressed as $0.42\cdot R_{L1}$, where $R_{L1}$ is the distance to the inner Lagrangian point $L1$. The radius uncertainties obtained from light curve fitting procedure are much larger than the expected deformation, therefore, negligible ellipsoidal modulations are expected.

We also have blocked the eclipses and fitted a sine function to the baseline modulation present in the folded light curve before performing the analysis with the JKTEBOP code. The modeled sine function was then added to the LC model and it was kept fixed. The new model presented better fitted residuals, nevertheless, the best-fit parameters did not change significantly and presented values within the estimated errors. Therefore, we adopted the simpler analysis in order to avoid unnecessary modifications of the folded light curve before the light curve synthesis procedure.

The errors given with the best-model parameters by {\it task $3$} are only formal errors and they are considered too optimistic. The errors obtained from this task are usually underestimated when one or more parameters are kept fixed during the fitting procedure, because correlations between parameters are discarded \citep{Torres00}. Thus, we have performed a Markov chain Monte Carlo (MCMC) simulation to obtain a more reliable error analysis (JKTEBOP {\it task $8$}), which included the photometric errors. By adding a Gaussian noise to the best-fitting model, the MCMC simulation provides the uncertainties from the distribution of best-fit solutions after many trials. We used 10000 chains to obtain a 1$\sigma$ uncertainty for the best-model parameters, which corresponds to the $68\%$ quartile half-width of the parameter distribution.

The final parameters derived for all five LMEBs are presented in table \ref{tab_res}. We have determined stellar masses between $0.244$ and $0.717$ $M_{\odot}$ and radii from $0.421$ to $0.670$ $R_{\odot}$. Our detached binaries are in very close orbits, with small orbital separations of $2.872 \leq a \leq 6.692$ $R_{\odot}$ and short periods of $0.59 \leq P_{\rm orb} \leq 1.72$ d, approximately.

%The best-fit model obtained for the EB 19g-2-08064 gave unusually small radius for the secondary component. 
%Some conditions together could have compromised this model, such as the limited precision of the light curve and the effective temperature derived spectroscopically. The primary and secondary eclipses shown in the EB 19g-2-08064 light curve have similar depths indicating that the binary components should have similar masses and therefore similar temperatures. This is supported by the radial velocity data, from which reliable masses of $0.71$ and $0.64$ $M_{\odot}$ could be found for the primary and the secondary components, respectively. Nevertheless, the spectral analysis suggests a $T_{\rm eff}$ of only $3100$ K for the secondary component. It is possible that flux calibration problems could have led to such discrepant temperature. The LC fitting converge to an unreasonable small secondary radius of $0.46$ $R_{\odot}$. 
%Given those inconsistencies, the estimated values for $R_{\rm 1}$ and $R_{\rm 2}$ -- and related quantities -- for the binary 19g-2-08064 should be taken with caution. Nevertheless, these values are listed as guidance for future analysis. 

The primary and secondary eclipses shown in the EB 19g-2-08064 light curve (see fig.\ref{figmodelfit19hr}) have similar depths indicating that the binary components should have similar masses and therefore similar temperatures. This is supported by the radial velocity data, from which reliable masses of $0.71$ and $0.64$ $M_{\odot}$ could be found for the primary and the secondary components, respectively. However, the spectral analysis suggested a $T_{\rm eff}$ of only $3100$ K for the secondary component (see sect. \ref{lowres}). It is possible that flux calibration problems could have led to such discrepant temperature.

The JKTABSDIM code \footnote{http://www.astro.keele.ac.uk/jkt/codes/jktabsdim.html.} \citep{Southworth05}, another resource associated with JKTEBOP, was used to obtain distance estimates for the binary systems. This procedure takes the results from JKTEBOP as inputs and accounts for a carefully performed error propagation. Among other quantities, JKTABSDIM calculates distances with a semi-empirical method based on the surface brightness-$T_{\rm eff}$ relation by \citet{Kervella04}, as described in \citet{Southworth05}, and which is suitable for main-sequence stars with effective temperatures from $3600$ to $10000$ K. Despite this temperature range, \citet{Kervella04} have presented indicators that this method is also valid in the infrared for dwarfs with temperatures as low as $\sim$$3000$ K \citep[][fig. 3]{Kervella04}.

%\begin{table}
%\begin{center}
%\caption{Interstellar extinction estimated for the studied LMEB systems. $A_{\rm V}$ was obtained by the relation $A_{\rm V} / {\rm E(B-V)} = 3.1$. All values are given in magnitude.}
%\label{tab_extinct}
%\begin{tabular}{lcc}
%%    \toprule
%\hline \hline \\ [-3ex]
%Binary  & ${\rm E(B-V)}$ & $A_{\rm V}$ \\
%%[0.5ex] 
%\hline\\  [-2ex]
%
%17e-3-02003 & 0.1601 $\pm$ 0.0011 & 0.4962 \\
%17h-4-01429 & 0.1270 $\pm$ 0.0035 & 0.3860 \\
%19c-3-08647 & 0.2041 $\pm$ 0.0119 & 0.6450 \\
%19f-4-05194 & 0.1263 $\pm$ 0.0047 & 0.3900 \\ 
%19g-2-08064 & 0.1572 $\pm$ 0.0106 & 0.4779 \\
%
%\hline %\hline
%\end{tabular}
%\end{center}
%\end{table}

The interstellar extinction has a strong effect at shorter wavelengths and may be significant at the near-infrared for the Galactic latitudes of our fields. Thus, we decided to perform an approximate reddening correction. 
The extinction was estimated for all 5 LMEBs by adopting the 3D extinction model by \citet{Arenou92}, which was computed with the help of their online service\footnote{Available at http://wwwhip.obspm.fr/cgi-bin/afm.}, and was considered for the distance estimation by the JKTABSDIM code. 
The computed distances range from $571$ to $1400$ pc and are presented in table \ref{tab_res}.

As a summary of the obtained results, we have characterized completely all 5 LMEBs, where the derived stellar masses lie between $0.24$ and $0.72$ $M_{\odot}$ and radii between $0.42$ to $0.67$ $R_{\odot}$, with orbital separations ranging from $2.87$ to $6.69$ $R_{\odot}$.

\section{Discussion}\label{discus}

Two of the studied LMEB systems, 17h-4-01429 and 19f-4-05194, were well-characterized where the radii were determined with uncertainties of around only 1\%. For three of our LMEBs, the WTS photometric data present a clear limitation on the accuracy of the measurements. Non-sequential time series with long time lapses between sampled cycles can be an issue when dealing with light curve modeling. This is noticeable through the visible dispersion of the data points in out-of-eclipse portions of the phase-folded light curves (figs. \ref{figmodelfit17hr} and \ref{figmodelfit19hr}). Such dispersion in phase-folded eclipses is reflected on the estimated error bars of the model-fit parameters, where the radii of both components are derived with larger uncertainties. 
This is the case of 17e-3-02003, 19c-3-08647, and 19g-2-08064, %\footnote{As discussed earlier in sect. \ref{sec:res}, the results obtained for the binary 19g-2-08064 should be taken with caution.}, 
where the dispersion present in their folded light curves is not ideal. The variability seen in the photometric data could be intrinsic to the objects. Phenomena like stellar spots, which could not be properly modeled with the data at hand, may differently affect in- and out-of-eclipse levels. Unfortunately, for these three binaries, additional multi-wavelength photometric follow up are needed in order to get more accurate radii with discriminating errors. 
%It is worth to mention, though, that error bars comparable to the ones derived for the radii of these 3 EBs ($\sim$15\%) are commonly obtained by studies that aim to characterize stellar physical parameters.

Nevertheless, despite other limitations on the spectroscopic data such as low SNR and incomplete phase coverage, the stellar masses of our whole sample (10 late-type stars) were derived with a very good precision. We obtained masses with errors of only 2.90\% and 5.64\% for the better and the worst cases, where the later was obtained for the least massive secondary component in our sample.

\subsection{The mass-radius diagram and the radius anomaly problem}\label{Ranomaly}

\begin{figure*}
\begin{center}
%\setcaptionmargin{1cm}
%\includegraphics[width=1.3 \columnwidth, angle=90]{plot_MRdiagram_mod.png}
%\includegraphics[width=1.8 \columnwidth, angle=90]{plot_MRdiagram_paper.png}
\includegraphics[width=1.8 \columnwidth, angle=0]{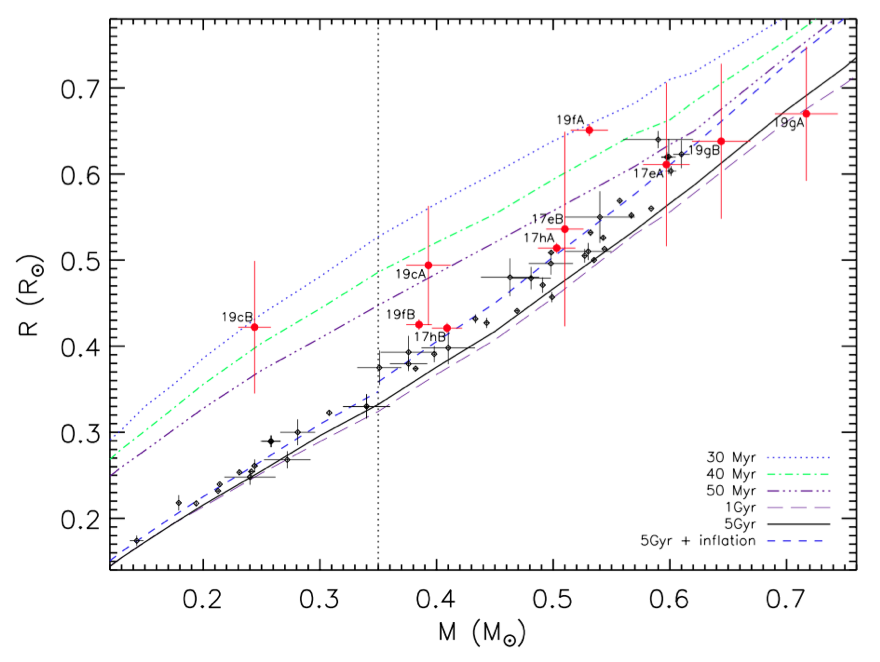}\caption{Mass-radius diagram. The isochrones of 30, 40, and 50 Myr and for 1 and 5 Gyr are the standard models from \citet{Baraffe98} ($[M/H]=0$, $Y=0.275$, $L_{mix} = H_{P}$). The additional 5 Gyr inflated evolutionary model was determined by \citet{Knigge11}. The LMEBs from the present work are represented by red filled circles and those gathered from the literature (with errors of less than 6\%) are illustrated as black small diamonds. Note that the EB components are presented here with shortened names for clarity of the figure.} 
\label{MRdiagram}
\end{center}
\end{figure*}

We have displayed our results on a mass-radius diagram along with selected well-studied low-mass EBs from the literature, which are shown in figure \ref{MRdiagram}. In this plot, the LMEBs from our sample are represented by red filled circles. The low-mass binaries from the literature are shown as black small diamonds and they are listed in the Appendix (Table \ref{tab_lit}).

Several isochrones are also plotted, for 30, 40, and 50 Myr, and for 1 and 5 Gyr. These isochrones are the standard models from \citet{Baraffe98}, calculated for a solar metallicity ($[M/H]=0$), a helium abundance of $Y=0.275$, and with a convective mixing length equal to the scale height ($L_{mix} = H_{P}$). 
An additional isochrone for 5 Gyr is illustrated by a blue dashed line that considers a radius inflation on the model of 4.5 and 7.9\% for different mass intervals of $M \leq 0.35$ $M_{\odot}$ and $M > 0.35$ $M_{\odot}$, respectively. These values were determined by \citet{Knigge11} by separating their sample in two groups: fully-convective stars with $M \leq 0.35$ $M_{\odot}$ and partially radiative stars with $M > 0.35$ $M_{\odot}$. This $0.35$ $M_{\odot}$ mass limit is represented in figure \ref{MRdiagram} as a vertical dotted line.

Our two well-characterized LMEB systems, 17h-4-01429 and 19f-4-05194, are placed in the region of partially radiative stars and seem to have inflated radii (see fig. \ref{MRdiagram}). The 17h-4-01429 system for example follows the 5 Gyr isochrone modified for a radius inflation of 7.9\%, where the primary 17h-4-01429 A and the secondary 17h-4-01429 B components are found just above this curve. In fact, we measured a higher radius inflation of 9.4 and 9.9\% for the primary and the secondary, respectively, when the obtained radii are compared to the 5 Gyr model-isochrone (black solid line, with no inflation). The primary and secondary components are presented in fig. \ref{MRdiagram} with shortened names (17hA and 17hB) for clarity of the figure.

The 19f-4-05194 system is the most interesting binary in our sample for being highly inflated when compared to the 5 Gyr model-isochrone, according to our results, where we estimated 31.1 and 17.2\% of radius inflation for the primary and secondary components, 19f-4-05194 A (19fA) and 19f-4-05194 B (19fB), respectively. \citet{Huelamo09} have found an EB system, HIP 96515 A, where the primary component is also more inflated than the secondary with a radius inflation of around 15 and 10\%, respectively, in comparison to evolutionary models. This system is part of a triple system with the white dwarf HIP 96515 B. According to \citet{Huelamo09}, the EB HIP 96515 A shows a fast rotation period and strong X-ray emission, what could be interpreted as signs of enhancement of the magnetic activity. 
Recently, another short-period EB KELTJ041621-620046 was found to have such radius anomaly, where the primary and the secondary components in this system are 28 and 17\% inflated, respectively \citep{Lubin17}. Nevertheless, the KELTJ041621-620046 system should be further characterized with more precision to confirm the observed large inflation, as the derived mass errors are of the order of $\sim$11\%.

Despite the uncertainty on the derived parameters, the LMEB 17e-3-02003 also seems to follow the same global inflation trend for partially convective stars (see fig. \ref{MRdiagram}). In this case, the primary and the secondary components (17eA and 17eB) present an inflation of 8.6 and 12.5\%, respectively.

The LMEB 19c-3-08647 is another interesting case that deserves further investigation. These objects compose the less massive binary and yet the most inflated components in our sample, with 33.6 and 68.6\% of radius inflation for the primary 19c-3-08647 A (19cA) and the secondary 19c-3-08647 B (19cB), respectively, when compared to the expected radii for a 5 Gyr model. Nevertheless, this binary in particular is placed closer to isochrones of younger stars, between the 30 and 50 Myr evolutionary models shown in the mass-radius diagram. If this LMEB is in fact a young system, with an age of around 40 Myr (green dot-dashed line, fig. \ref{MRdiagram}), it would justify then such inflated radii for both components. 
It is not possible though to ensure the age of the system based only on isochrones. 
Therefore, additional high-resolution spectroscopic data with good SNR are needed to look for other age tracers.

The most massive binary in our sample, 19g-2-08064 ($M_{pri}$=$0.71$ and $M_{sec}$=$0.64$ $M_{\odot}$), seems to be the less inflated system. The secondary 19g-2-08064 B (19gB) presents an inflation of around 6.0\% when compared to the 1 Gyr model-isochrone (purple long-dashed line, fig. \ref{MRdiagram}) and the primary 19g-2-08064 A (19gA) does not show any inflation. When compare to the 5 Gyr model-isochrone, 19gB is only 4.1\% inflated.

As a general remark, the great majority of the LMEBs in our sample has an estimated radius far from the predicted values according to evolutionary models, where the studied low-mass stars seem to be significantly inflated. In particular, the components with derived masses of $M < 0.6$ $M_{\odot}$ present a radius inflation of $\sim$$9\%$ or more. This result supports a global trend of inflation for partially-radiative stars ($M > 0.35$ $M_{\odot}$), as discussed by \citet{Knigge11}, for low-mass stars in close detached low-mass binary systems, with a possible exception for the components of the EB that could be a young system.

%As a general remark, none of the fully characterized LMEBs in our sample has an estimated radius close to the predicted values according to evolutionary models. All studied low-mass stars seem to be inflated of about $9\%$ or more, with a possible exception for the components of the EB that could be a young system. This result supports a global trend of inflation for partially-radiative stars ($M > 0.35$ $M_{\odot}$), as discussed by \citet{Knigge11}, for low-mass stars in close detached low-mass binary systems.

As mentioned briefly at the beginning of this paper, the radius inflation problem is not a new issue in the field and a few theories have been discussed for some time in the literature as attempts of finding the processes that are causing this radius anomaly. For example, \citet{Chabrier07} suggested that the observed inflated radii of low-mass stars in binary systems maybe be caused by a less efficient heat transport. An enhanced magnetic activity was pointed out then as the possible causer of the radius inflation \citep{Torres06,Chabrier07}.

Indeed, considering that these low-mass stars are in close but still detached systems with short orbital periods (of around 2 days), they should be synchronized and in circular orbits due to tidal effects \citep{Zahn77}. This synchronization, or tidal-locking, causes an enhancement of the magnetic activity that becomes an important process that should be taken into account. 
According to \citet{Chabrier07}, the magnetic activity may reduce the convective efficiency in partially-radiative stars -- where a smaller convective mixing length leaves a larger amount of heat to be transported via radiation --, which results in a cooler and more inflated star in order to keep thermal equilibrium. This hypothesis is sustained in the literature by observations of H${\rm \alpha}$ and X-ray emission from one or both of the binary components. It is worth to emphasize then that all LMEB systems in our sample have spectra with chromospheric H${\rm \alpha}$ emission from both components of the system.

As a counterexample, the recent study of the chromospherically active LMEB T-Cyg1-12664 by \citet{Han17} shows that neither component of the binary is inflated with respect to models considering their new derived stellar physical parameters, with radii of $0.92$ and $0.47$ $R_{\odot}$ and masses of $0.91$ and $0.50$ $M_{\odot}$ for the primary and secondary components, respectively. However, this EB's primary component is placed in a region of more massive primary stars than the objects discussed in this work ($M < 0.7$ $M_{\odot}$).

Moreover, the chromospheric activity can also be correlated with rotation \citep{Bouvier90}, thus, fast rotators may have inflated radii. In fact, fast rotators during the early pre-MS phase may have their radii enlarged by strong magnetic fields \citep[][and references therein]{Somers14,Barrado16}.

It is not of the scope of this work to unveil the causes of the radius inflation but to provide additional measurements in the intriguing scenario of inflated components in LMEBs. Given the restricted known sample, every single new well-characterized low-mass system is significant on the pursuit of the causes of the radius anomaly. 
Then, these systems add to the increasing sample of low-mass stellar radii that are not well-reproduced by stellar models and they highlight the need to understand the magnetic activity and physical state of small stars. 
Space missions, like TESS for example, will provide many such systems to perform high-precision radius measurements to tightly constrain low-mass stellar evolution models.

\section{Conclusions}

The performed analysis was dedicated to the characterization of 5 new short-period low-mass eclipsing binaries from the WFCAM Transit Survey, with periods of less than 2 days.

Low- and intermediate-resolution spectroscopic data were acquired in order to obtain stellar atmospheric parameters and to measure radial velocity shifts via cross-correlation. The low-resolution spectra have confirmed the low-mass nature of the binaries, with effective temperatures ranging from $3000$ to $4400$ K. The higher resolution spectra have revealed double-peaked H${\rm \alpha}$ emission, indicating the presence of chromospheric activity in both components of the systems.

The WFCAM J-band light curves and the RV shifts were modeled together with the JKTEBOP code. From the best-model fit, we derived stellar masses between $0.24$ and $0.72$ $M_{\odot}$ and radii from $0.42$ to $0.67$ $R_{\odot}$. The completely characterized binaries are in very close orbits, with orbital separations of $2.87 \leq a \leq 6.69$ $R_{\odot}$ and short periods of $0.59 \leq P_{\rm orb} \leq 1.72$ d, approximately. We also estimated that these detached systems are at distances from $0.5$ to $1.4$ kpc.

The stellar masses of our 10 late-type stars were determined with a precision of 6\% or better.
Two LMEBs, 17h-4-01429 and 19f-4-05194, were well characterized with the analyzed data, where we obtained radius errors of around 1\%. For the EBs 17e-3-02003, 19c-3-08647 and 19g-2-08064, the radii were determined with much higher uncertainties due to the limited precision of the WTS photometric data, but still indicate notable inflation above the radii expected from standard stellar models. 
%However, error estimates of around 15\% are common in the literature for the determination of stellar physical parameters. 

%The EBs investigated in this work follow the global trend of inflation for partially-radiative stars proposed by \citet{Knigge11}, and supports the idea of the radius anomaly as a real phenomenon of low-mass star in close detached low-mass binary systems.

\section*{Acknowledgements}

PC and MD would like to thank Dr. Hern\'an Garrido for all fruitful discussions. The authors would also like to thank Dr. David Pinfield, and the WTS Consortium. 
This research has been funded by Brazilian FAPESP grant 2015/18496-8. MD thanks CNPq funding under grant \#305657. This research has been funded by Spanish grant ESP2015-65712-C5-1-R.
PC, DB, BS and SH have received support from the RoPACS network during this research, a Marie Curie Initial Training Network funded by the European Commissions Seventh Framework Programme FP7-PEOPLE-2007-1-1-ITN. 
This article is based on data collected under Service Time program at the Calar Alto Observatory, the German-Spanish Astronomical Center, Calar Alto, jointly operated by the Max-Planck-Institut f\"ur Astronomie Heidelberg and the Instituto de Astrof\'isica de Andaluc\'ia (CSIC). We are very grateful to the CAHA staff for the superb quality of the observations. We also thank the Calar Alto Observatory for the allocation of director's discretionary time to this program. 
This publication makes use of VOSA, developed under the Spanish Virtual Observatory project supported from the Spanish MICINN through grant AyA2011-24052. This work has made use of the ALADIN interactive sky atlas and the SIMBAD database, operated at CDS, Strasbourg, France, and of NASA's Astrophysics Data System Bibliographic Services. %This research has made use of the NASA/IPAC Infrared Science Archive, which is operated by the Jet Propulsion Laboratory, California Institute of Technology, under contract with the National Aeronautics and Space Administration.

%%%%%%%%%%%%%%%%%%%%%%%%%%%%%%%%%%%%%%%%%%%%%%%%%%



%\bibliographystyle{mnras}


%%%%%%%%%%%%%%%%%%%%%%%%%%%%%%%%%%%%%%%%%%%%%%%%%%

%%%%%%%%%%%%%%%%% APPENDICES %%%%%%%%%%%%%%%%%%%%%

\appendix

\section{Low-mass binaries from the literature}\label{appen}

Table \ref{tab_lit} shows the list of low-mass eclipsing binaries found in the literature that are illustrated in figure \ref{MRdiagram}. The components of these systems are low-mass stars, with $R \leq 0.7$ $R_{\odot}$ and $M \leq 0.7$ $M_{\odot}$. Only EB systems where both components are low-mass stars where considered. These systems also have orbital periods of less than 5 days. It is important to mention that these are selected as well-characterized systems, with errors of less than $6\%$ on the radii and masses. Only three stars listed in table \ref{tab_lit} have mass errors greater than $6\%$ (but less than $10\%$). Nevertheless, they were kept in the sample because they present radii defined with good precision.

%\begin{landscape}
\begin{table*}
\begin{center}
\caption{Low-mass binaries from the literature with $P_{\rm orb} \leq 5$ days. The following binaries present radius errors of less than 6\%. The errors in mass are typically smaller than 6\%, with a few exceptions.
}
\label{tab_lit}
\begin{tabular}{lcccc}
\hline \hline \\ [-3ex]
%Star  & Period & Mass & Radius & $T_{\rm eff}$ & Reference \\
%	  & ($d$) & ($M\odot$) & ($R\odot$) & ($K$) &  \\
Star  & Period & Mass & Radius & Reference \\
	  & ($d$) & ($M_{\odot}$) & ($R_{\odot}$) &  \\

%[0.5ex] 
\hline\\  [-2ex]

NSVS01031772 A & 0.368 & 0.5428 $\pm$ 0.0028 & 0.5260 $\pm$ 0.0028 & \cite{Southworth15}$^{\rm *}$ \\ % & 3614.1 $\pm$ 67.2 & 1 \\
NSVS01031772 B & 0.368 & 0.4982 $\pm$ 0.0025 & 0.5087 $\pm$ 0.0031 & \cite{Southworth15}$^{\rm *}$ \\ % & 3515.6 $\pm$ 32.5 & 1 \\
SDSS-MEB-1 A & 0.407037 & 0.272 $\pm$ 0.020 & 0.268 $\pm$ 0.010 & \cite{Blake08} \\ % & 3320 $\pm$ 130 & 14 \\
SDSS-MEB-1 B & 0.407037 & 0.240 $\pm$ 0.022 & 0.248 $\pm$ 0.0090 & \cite{Blake08} \\ % & 3300 $\pm$ 130 & 14 \\
GU Boo A & 0.489 & 0.6100 $\pm$ 0.0071 & 0.6230 $\pm$ 0.0163 & \cite{Southworth15}$^{\rm *}$ \\ % & 3917.4 $\pm$ 128.3 & 1 \\
GU Boo B & 0.489 & 0.5990 $\pm$ 0.0061 & 0.6200 $\pm$ 0.0203 & \cite{Southworth15}$^{\rm *}$ \\ % & 3810.7 $\pm$ 133.9 & 1 \\
MG1-1819499 A & 0.6303135 & 0.557 $\pm$ 0.001 & 0.569 $\pm$ 0.002 & \cite{Kraus11} \\ % & 3690.0 $\pm$ 100.0 & 2 \\
MG1-1819499 B & 0.6303135 & 0.535 $\pm$ 0.001 & 0.500 $\pm$ 0.003 & \cite{Kraus11} \\ % & 3610.0 $\pm$ 100.0 & 2 \\
GJ 3236 A & 0.77126 & 0.376 $\pm$ 0.016 & 0.3795 $\pm$ 0.0084 & \cite{Irwin09} \\ % & 3312.0 $\pm$ 110.0 & 3 \\
GJ 3236 B & 0.77126 & 0.281 $\pm$ 0.015 & 0.300 $\pm$ 0.015 & \cite{Irwin09} \\ % & 3242.0 $\pm$ 108.0 & 3 \\
YY Gem A & 0.814 & 0.5974 $\pm$ 0.0047 & 0.6196 $\pm$ 0.0057 & \cite{Southworth15}$^{\rm *}$ \\ % & 3819.4 $\pm$ 98.0 & 1 \\
YY Gem B & 0.814 & 0.6009 $\pm$ 0.0047 & 0.6035 $\pm$ 0.0057 & \cite{Southworth15}$^{\rm *}$ \\ % & 3819.4 $\pm$ 98.0 & 1 \\
MG1-116309 A & 0.8271425 & 0.567 $\pm$ 0.002 & 0.552 $\pm$ 0.004 & \cite{Kraus11} \\ % & 3917.4 $\pm$ 100.5 & 2 \\
MG1-116309 B & 0.8271425 & 0.532 $\pm$ 0.002 & 0.532 $\pm$ 0.004 & \cite{Kraus11} \\ % & 3810.7 $\pm$ 97.8 & 2 \\
CM Dra A & 1.268 & 0.2310 $\pm$ 0.0009 & 0.2534 $\pm$ 0.0019 & \cite{Southworth15}$^{\rm *}$ \\ % & 3133.3 $\pm$ 73.0 & 1 \\
CM Dra B & 1.268 & 0.2141 $\pm$ 0.0009 & 0.2396 $\pm$ 0.0015 & \cite{Southworth15}$^{\rm *}$ \\ % & 3118.9 $\pm$ 102.2 & 1 \\
19b-2-01387 A & 1.49851768 & 0.498 $\pm$ 0.019 & 0.496 $\pm$ 0.013 & \cite{Birkby12} \\ % & 3498 $\pm$ 100 & 10 \\
19b-2-01387 B & 1.49851768 & 0.481 $\pm$ 0.017 & 0.479 $\pm$ 0.013 & \cite{Birkby12} \\ % & 3436 $\pm$ 100 & 10 \\
MG1-506664 A & 1.5484492 & 0.584 $\pm$ 0.002 & 0.560 $\pm$ 0.001 & \cite{Kraus11} \\ % & 3732.5 $\pm$ 104.6 & 2 \\
MG1-506664 B & 1.5484492 & 0.544 $\pm$ 0.002 & 0.513 $\pm$ 0.001 & \cite{Kraus11} \\ % & 3614.1 $\pm$ 101.3 & 2 \\
MG1-78457 A & 1.5862046 & 0.5270 $\pm$ 0.0019 & 0.505 $\pm$ 0.008 & \cite{Kraus11} \\ % & 3326.6 $\pm$ 101.1 & 2 \\
MG1-78457 B & 1.5862046 & 0.491 $\pm$ 0.002 & 0.471 $\pm$ 0.009 & \cite{Kraus11} \\ % & 3273.4 $\pm$ 99.5 & 2 \\
LP133−373 A & 1.6279866 & 0.34 $\pm$ 0.02 & 0.330 $\pm$ 0.014 & \cite{Vaccaro07} \\ % & 3144.0 $\pm$ 206.0 & 4 \\
LP133−373 B & 1.6279866 & 0.34 $\pm$ 0.02 & 0.330 $\pm$ 0.014 & \cite{Vaccaro07} \\ % & 3058.0 $\pm$ 195.0 & 4 \\
MG1-646680 A & 1.6375302 & 0.499 $\pm$ 0.002 & 0.457 $\pm$ 0.006 & \cite{Kraus11} \\ % & 3732.5 $\pm$ 51.9 & 2 \\
MG1-646680 B & 1.6375302 & 0.443	0.002 & 0.427 $\pm$ 0.006 & \cite{Kraus11} \\ % & 3630.8 $\pm$ 50.5 & 2 \\
19e-3-08413 A & 1.67343720 & 0.463 $\pm$ 0.025 & 0.480 $\pm$ 0.022 & \cite{Birkby12} \\ % & 3506 $\pm$ 140 & 10 \\
19e-3-08413 B & 1.67343720 & 0.351 $\pm$ 0.019 & 0.375 $\pm$ 0.020 & \cite{Birkby12} \\ % & 3338 $\pm$ 140 & 10 \\
MG1-2056316 A & 1.7228208 & 0.4690 $\pm$ 0.0021 & 0.441 $\pm$ 0.002 & \cite{Kraus11} \\ % & 3459.4 $\pm$ 179.8 & 2 \\
MG1-2056316 B & 1.7228208 & 0.382 $\pm$ 0.002 & 0.374 $\pm$ 0.002 & \cite{Kraus11} \\ % & 3318.9 $\pm$ 172.5 & 2 \\
KOI126 B & 1.76713 & 0.2413 $\pm$ 0.0030 & 0.2543 $\pm$ 0.0014 & \cite{Carter11} \\ % & 0.0 $\pm$ 0.0 & 5 \\
KOI126 C & 1.76713 & 0.2127 $\pm$ 0.0026 & 0.2318 $\pm$ 0.0013 & \cite{Carter11} \\ % & 0.0 $\pm$ 0.0 & 5 \\
HIP 96515 Aa & 2.3456 & 0.59 $\pm$ 0.03 & 0.64 $\pm$ 0.01 & \cite{Huelamo09} \\ % & 3724.0 $\pm$ 154.0 & 6 \\
HIP 96515 Ab & 2.3456 & 0.54 $\pm$ 0.03 & 0.55 $\pm$ 0.03 & \cite{Huelamo09} \\ % & 3589.0 $\pm$ 157.0 & 6 \\
19g-4-02069 A & 2.44178 & 0.53 $\pm$ 0.02 & 0.51 $\pm$ 0.01 & \cite{Nefs13} \\ % & 3300 $\pm$ 140 & 12 \\
19g-4-02069 B & 2.44178 & 0.143 $\pm$ 0.006 & 0.174 $\pm$ 0.006 & \cite{Nefs13} \\ % & 2950 $\pm$ 140 & 12 \\
CU Cnc A & 2.771 & 0.4333 $\pm$ 0.0017 & 0.4317 $\pm$ 0.0052 & \cite{Southworth15}$^{\rm *}$ \\ % & 3162.3 $\pm$ 156.7 & 1 \\
CU Cnc B & 2.771 & 0.3980 $\pm$ 0.0014 & 0.3908 $\pm$ 0.0095 & \cite{Southworth15}$^{\rm *}$ \\ % & 3126.1 $\pm$ 154.9 & 1 \\
1RXSJ154727 A & 3.5500184 & 0.2576 $\pm$ 0.0085 & 0.2895 $\pm$ 0.0068 & \cite{Hartman11} \\ % & 0.0 $\pm$ 0.0 & 7 \\
1RXSJ154727 B & 3.5500184 & 0.2585 $\pm$ 0.0080 & 0.2895 $\pm$ 0.0068 & \cite{Hartman11} \\ % & 0.0 $\pm$ 0.0 & 7 \\
HATS557-027 A & 4.077017 & 0.244 $\pm$ 0.003 & 0.261$^{0.006}_{0.009}$ & \cite{Zhou15} \\ % & 3190 $\pm$ 100 & 11 \\
HATS557-027 B & 4.077017 & 0.179$^{0.002}_{0.001}$ & $0.218^{0.007}_{0.011}$ & \cite{Zhou15} \\ % & 2990 $\pm$ 110 & 11 \\
LP661-13 A & 4.7043512 & 0.30795 $\pm$ 0.00084 & 0.3226 $\pm$ 0.0033 & \cite{Dittmann17} \\ % & 0.0 $\pm$ 0.0 & 13 \\
LP661-13 B & 4.7043512 & 0.19400 $\pm$ 0.00034 & 0.2174 $\pm$ 0.0023 & \cite{Dittmann17} \\ % & 0.0 $\pm$ 0.0 & 13 \\
19c-3-01405 A & 4.9390945 & 0.410 $\pm$ 0.023 & 0.398 $\pm$ 0.019 & \cite{Birkby12} \\ % & 3309 $\pm$ 130 & 10 \\
19c-3-01405 B & 4.9390945 & 0.376 $\pm$ 0.024 & 0.393 $\pm$ 0.019 & \cite{Birkby12} \\ % & 3305 $\pm$ 130 & 10 \\
%LSPMJ1112 A & 41.03236 & 0.3946 $\pm$ 0.0023 & 0.3860 $\pm$ 0.005 & \cite{Irwin11} \\ % & 3061.0 $\pm$ 162.0 & 8 \\
%LSPMJ1112 B & 41.03236 & 0.2745 $\pm$ 0.0012 & 0.2978 $\pm$ 0.005 & \cite{Irwin11} \\ % & 2952.0 $\pm$ 163.0 & 8 \\
%Kepler-16 A & 41.079220 & 0.6897 $\pm$ 0.0035 & 0.6489 $\pm$ 0.0013 & \cite{Doyle11} \\ % & 4450 $\pm$ 150 & 9 \\
%Kepler-16 B & 41.079220 & 0.20255 $\pm$ 0.00066 & 0.22623 $\pm$ 0.00059 & \cite{Doyle11} \\ % & 0.0 $\pm$ 0.0 & 9 \\
%

\hline %\hline
\end{tabular}
\begin{list}{}{}
\item[]{\scriptsize{{\bf Note.} $^{\rm *}$ And references therein.}}
\end{list}
\end{center}
\end{table*}
%\end{landscape}


%%%%%%%%%%%%%%%%%%%%%%%%%%%%%%%%%%%%%%%%%%%%%%%%%%

% Don't change these lines
\bsp	% typesetting comment
\label{lastpage}
\end{document}